\documentclass[aps,floatfix,superscriptaddress,preprintnumbers,showpacs,showkeys]{revtex4}
\usepackage{amssymb}
\usepackage{amsmath}
\usepackage{amsfonts}
\usepackage{epsfig}
\usepackage{graphicx}
\usepackage{tabularx}
\usepackage{verbatim}
\newcommand{\eq}{\begin{eqnarray}}
\newcommand{\en}{\end{eqnarray}}

\begin{document}

\title{Effect of the $K\overline{K}$ and $\eta\eta$ channels and interference phenomena 
in the two-pion and $K\overline{K}$ transitions of charmonia and bottomonia}

\author{Yury S. Surovtsev}
\affiliation{Bogoliubov Laboratory of Theoretical Physics,
Joint Institute for Nuclear Research, 141980 Dubna, Russia}
\author{Petr Byd\v{z}ovsk\'y}
\affiliation{Nuclear Physics Institute of the AS CR, 25068 \v{R}e\v{z},
Czech Republic}
\author{Thomas Gutsche}
\affiliation{Institut f\"ur Theoretische Physik,
Universit\"at T\"ubingen,
Kepler Center for Astro and Particle Physics,
Auf der Morgenstelle 14, D-72076 T\"ubingen, Germany}
\author{Robert~Kami\'nski}
\affiliation{Institute of Nuclear Physics PAS, Cracow 31342, Poland}
\author{Valery E. Lyubovitskij}
\affiliation{Institut f\"ur Theoretische Physik,
Universit\"at T\"ubingen,
Kepler Center for Astro and Particle Physics,
Auf der Morgenstelle 14, D-72076 T\"ubingen, Germany}
\affiliation{Departamento de F\'\i sica y Centro Cient\'\i fico
Tecnol\'ogico de Valpara\'\i so-CCTVal, Universidad T\'ecnica
Federico Santa Mar\'\i a, Casilla 110-V, Valpara\'\i so, Chile}
\affiliation{Department of Physics, Tomsk State University,
634050 Tomsk, Russia}
\affiliation{Laboratory of Particle Physics, Tomsk Polytechnic University,
634050 Tomsk, Russia}
\author{Miroslav Nagy}
\affiliation{Institute of Physics, SAS, Bratislava 84511, Slovak Republic}

\date{\today}

\begin{abstract}

It is shown that the basic shape of dipion and $K\overline{K}$ mass spectra 
in decays $J/\psi\to\phi(\pi\pi,K\overline{K})$, $\psi(2S)\to J/\psi\,\pi\pi$, 
$Y(4260)\to J/\psi~\pi\pi$ and in the two-pion transitions of bottomonia states 
are explained by an unified mechanism based on the contribution of the 
$\pi\pi$, $K\overline{K}$ and $\eta\eta$ coupled channels including their 
interference. The role of the individual $f_0$ resonances in making up the shape 
of the dipion mass distributions in the charmonia and bottomonia decays is considered.

\end{abstract}

\pacs{11.55.Bq,11.80.Gw,12.39.Mk,14.40.Pq}

\keywords{coupled--channel formalism, meson--meson scattering,
heavy meson decays, scalar and pseudoscalar mesons}

\maketitle

\section{Introduction}
The last years achievements of the hadron spectroscopy are related mainly 
to heavy mesons including charmonia and bottomonia. Therefore, of course, 
there is a problem of studying structure of these mesons and their interaction 
for exploring nonperturbative QCD. There were expressed thoughts that for these 
aims the two-pion transitions of bottomonia  are suitable (see, e.g., 
\cite{Simonov-Veselov} and references therein). Clearly, first it is necessary 
to allow for the final-state interaction in decays of bottomonia. 
When studying processes $\Upsilon(mS)\to\Upsilon(nS)\pi\pi$, the $\pi\pi$ 
interaction should be considered with taking into account the coupled channels, 
whereas the final vector meson $\Upsilon(nS)$ remains a spectator.

In Refs.~\cite{SBGKLN-pr15_1,SBGKLN-pr15_2}, devoted to explanation of the 
two-pion transitions of bottomonia, we have shown that the basic shape of 
dipion mass spectra in the decays both of bottomonia and charmonia are explained 
by the unified mechanism which is based on our previous conclusions on wide 
resonances~\cite{SBKLN-prd12,SBLKN-jpgnpp14} and is related to contributions 
of the $\pi\pi$ and $K\overline{K}$ coupled channels including their interference
and allowing for our earlier results on the explanation of charmonia decays  
$J/\psi\to\phi(\pi\pi,K\overline{K})$ and 
$\psi(2S)\to J/\psi(\pi\pi)$~\cite{SBLKN-prd14,SBGLKN-npbps13}. 
In indicated decays pseudoscalar meson pairs are produced mainly in the 
scalar-isoscalar state. In Refs.~\cite{SBKLN-prd12,SBLKN-jpgnpp14} it was shown 
that correct parameters of the $f_0$ mesons cannot be obtained 
when studying only 
the $\pi\pi$ scattering. Allowance for the coupled $K\overline{K}$ channel and 
consideration of the corresponding experimental data improve situation. 
However, in order to extract more correct values of the $f_0$ parameters it is needed 
to take into account also the coupled $\eta\eta$ channel. Note that here talking must 
be about the combined analysis of data on the isoscalar S-wave processes 
$\pi\pi\to\pi\pi,K\overline{K},\eta\eta$, of accessible data on the charmonia decay 
processes --- $J/\psi\to\phi(\pi\pi,K\overline{K})$, $\psi(2S)\to J/\psi\pi\pi$ 
(Crystal Ball~\cite{Crystal_Ball_80}, DM2~\cite{DM2}, Mark~II~\cite{Mark_II}, 
Mark~III~\cite{Mark_III}, and BES~II~\cite{BESII} Collaborations) and of practically 
all available data on two-pion transitions of the $\Upsilon$ mesons from 
the ARGUS~\cite{Argus}, CLEO~\cite{CLEO}, CUSB~\cite{CUSB}, 
Crystal Ball~\cite{Crystal_Ball(85)}, Belle~\cite{Belle}, and 
{\it BABAR}~\cite{BaBar06} Collaborations --- 
$\Upsilon(mS)\to\Upsilon(nS)\pi\pi$ ($m>n$, $m=2,3,4,5,$ $n=1,2,3$).

When analyzing the charmonia and bottomonia decays with allowing for also the 
$\eta\eta$ channel, we need to include the $\pi\pi\to\eta\eta$ and 
$K\overline{K}\to\eta\eta$ amplitudes, respectively. In the former case the phase 
shift of $\pi\pi\to\eta\eta$ amplitude is unknown from data, in the latter
the $K\overline{K}\to\eta\eta$ amplitude is experimentally unknown entirely. 
However, thanks to the Le~Couteur--Newton relations~\cite{LeCou}, which represent 
all amplitudes of transitions between three coupled channels ($\pi\pi$, $K\overline{K}$ 
and $\eta\eta$) via one function -- the Jost matrix determinant, we know the 
model-independent part of the amplitude related to resonances. 
The only remaining important problem is the description of the background part. 
After solution of this problem we shall be able to predict the unknown indicated 
amplitudes.

Thus, the main aim of this investigation is to prolong the study of scalar meson 
properties analyzing jointly data on the isoscalar S-wave processes 
$\pi\pi\to\pi\pi,K\overline{K},\eta\eta$, on charmonia decays --- 
$J/\psi\to\phi(\pi\pi,K\overline{K})$, $\psi(2S)\to J/\psi\pi\pi$ and 
$Y(4260)\to J/\psi~\pi\pi$ and on the above-indicated two-pion transitions 
of bottomonia. This task is timely in view of that is very important. However, 
up to now it is not completely carried out. E.g., analyzing the multichannel 
$\pi\pi$ scattering and the decays $J/\psi\to\phi(\pi\pi,K\overline{K})$ with 
the data of Mark~III and DM2 Collaborations in the framework of our approach 
based on analyticity and unitarity and with using an uniformization procedure, 
we have obtained parameters of the $f_0(600)$ and $f_0(1500)$ which differ 
considerably from results of analyses based on some other methods (mainly 
those based on the dispersion relations and Breit -- Wigner 
approaches)~\cite{SBL-prd12,PDG-16}. Moreover, it was found that the data admit 
two sets of parameters of $f_0(500)$ with a mass relatively near to the 
$\rho$-meson mass, and with the total widths either $\approx 600$ or 
$\approx 930$~MeV. Addition to the combined analysis the BES~II data on $
J/\psi\to\phi\pi\pi$ has given the important result choosing surely from 
two solutions for the $f_0(500)$ the one with the larger width~\cite{SBLKN-prd14}. 
When expanding the analysis via adding the data on decays 
$\psi(2S)\to J/\psi\,\pi\pi$ and on the two-pion transitions of bottomonia, 
the satisfactory description did not require the alteration of the $f_0$ parameters, 
thus confirming our earlier conclusions about the scalar-isoscalar mesons. 
Also there was obtained the interesting and unified explanation of dipion mass 
spectra for the indicated charmonia and bottomonia 
decays~\cite{SBGKLN-pr15_1,SBGKLN-pr15_2}. 
E.g., we have showed that the experimentally observed interesting behavior of the 
$\pi\pi$ spectra of the $\Upsilon$-family decays, beginning from the second radial 
excitation and higher, --- a bell-shaped form in the near-$\pi\pi$-threshold region, 
smooth dips about 0.6~GeV in the $\Upsilon(4S,5S)\to\Upsilon(1S) \pi^+ \pi^-$, 
about 0.45~GeV in the $\Upsilon(4S,5S)\to\Upsilon(2S) \pi^+ \pi^-$, 
and about 0.7~GeV in the $\Upsilon(3S)\to\Upsilon(1S)(\pi^+\pi^-,\pi^0\pi^0)$, 
and also sharp dips about 1~GeV in the $\Upsilon(4S,5S)\to\Upsilon(1S) \pi^+ \pi^-$ --- 
is explained by the interference between the $\pi\pi$ scattering and 
$K\overline{K}\to\pi\pi$ contributions to the final states of these decays 
(by the constructive one in the near-$\pi\pi$-threshold region and by the destructive 
one in the dip regions).

Clearly, the allowance for effect of the $\eta\eta$ channel in the indicated 
two-pion transitions (as of the $\pi\pi$ and $K\overline{K}$ channels) not 
only kinematically (i.e. taking account the channel threshold via the uniformizing 
variable) and also by adding the $\pi\pi\to\eta\eta$ amplitude in the corresponding 
formulas for the decays permit us to extend description of the $\pi\pi$ spectra of 
relevant decays above the $\eta\eta$ threshold. Besides specifying the decay 
parameters, we are going to clarify a role of individual resonances.

\section{The effect of multichannel $\pi\pi$ scattering in decays of the 
$\psi$- and $\Upsilon$-meson families}

Considering multichannel $\pi\pi$ scattering, we shall deal with the 3-channel case,  
i.e. with the reactions $\pi\pi\to\pi\pi,K\overline{K},\eta\eta$, because it was 
shown~\cite{SBLKN-jpgnpp14} that this is a minimal number of coupled channels needed 
for obtaining correct values of $f_0$-resonance parameters. 
When performing our combined analysis data for the multichannel $\pi\pi$ scattering 
were taken from many papers (see Refs. in our paper~\cite{SBLKN-prd14}).
For the decay $J/\psi\to\phi\pi^+\pi^-$ data were taken from 
Mark III, DM2 and BES II Collaborations;
for $\psi(2S)\to J/\psi(\pi^+\pi^-~{\rm and}~\pi^0\pi^0)$ --- 
from Mark~II and Crystal Ball(80) (see Refs. also in~\cite{SBLKN-prd14}).
For $\Upsilon(2S)\to\Upsilon(1S)(\pi^+\pi^-~{\rm and}~\pi^0\pi^0)$ data were used 
from ARGUS~\cite{Argus}, CLEO~\cite{CLEO}, CUSB~\cite{CUSB}, and 
Crystal Ball~\cite{Crystal_Ball(85)} Collaborations; for
$\Upsilon(3S)\to\Upsilon(1S)(\pi^+\pi^-,\pi^0\pi^0)$ and 
$\Upsilon(3S)\to\Upsilon(2S)(\pi^+\pi^-,\pi^0\pi^0)$ --- 
from CLEO \cite{CLEO(94),CLEO07}; for $\Upsilon(4S)\to\Upsilon(1S,2S)\pi^+\pi^-$ --- 
from {\it BABAR} \cite{BaBar06} and Belle \cite{Belle}; 
for $\Upsilon(5S)\to\Upsilon(1S,2S,3S)(\pi^+\pi^-,\pi^0\pi^0)$ --- 
from Belle Collaboration~\cite{Belle, Belle2013}.

The used formalism for calculating the dimeson mass distributions in the quarkonia 
decays is analogous to the one proposed in Ref.~\cite{MP-prd93} for the decays 
$J/\psi\to\phi(\pi\pi, K\overline{K})$ and $V^{\prime}\to V\pi\pi$ ($V=\psi,\Upsilon$) 
but with allowing for also amplitudes of transitions between the $\pi\pi$, 
$K\overline{K}$ and $\eta\eta$ channels in decay formulas. 
There was assumed that 
the mesons pairs in the final state have zero isospin and spin. 
Only these pairs of mesons undergo final state interactions whereas 
the final $\Upsilon(nS)$ meson ($n<m$) 
remains a spectator. The amplitudes of decays are related with the scattering amplitudes
$T_{ij}$ $(i,j=1-\pi\pi,2-K\overline{K},3-\eta\eta)$ as follows
\begin{eqnarray}
&&F\bigl(J/\psi\to\phi\pi\pi\bigr)=\frac{1}{\sqrt{3}}~\bigl[c_1(s)T_{11}+
\Bigl(\frac{\alpha_2}{s-\beta_2}+c_2(s)\Bigr)T_{12}+c_3(s)T_{13}\bigr],\\
&&F\bigl(J/\psi\to\phi K\overline{K}\bigr)=
\frac{1}{\sqrt{2}}~\bigl[c_1(s)T_{21}+c_2(s)T_{22}+c_3(s)T_{23}\bigr],\\
&&F\bigl(\psi(2S)\to\psi(1S)\pi\pi\bigr)=\frac{1}{\sqrt{3}}~\bigl[d_1(s)T_{11}
+d_2(s)T_{12}+d_3(s)T_{13}\bigr],\\
&&F\bigl(\Upsilon(mS)\to\Upsilon(nS)\pi\pi\bigr) = 
\frac{1}{\sqrt{3}}~\bigl[e_1^{(mn)}T_{11}+ e_2^{(mn)}T_{12}+ e_3^{(mn)}T_{13}\bigr],\\ 
&&~~~~~~~~~~~~~~~~~~m>n,~ m=2,3,4,5,~ n=1,2,3\nonumber
\end{eqnarray}
where $c_i=\gamma_{i0}+\gamma_{i1}s$, $d_i=\delta_{i0}+\delta_{i1}s$ and 
$e_i^{(mn)}=\rho_{i0}^{(mn)}+\rho_{i1}^{(mn)}s$; indices $m$ and $n$ 
correspond to $\Upsilon(mS)$ and $\Upsilon(nS)$, respectively. 
The free parameters $\alpha_2$, $\beta_2$, $\gamma_{i0}$, $\gamma_{i1}$, 
$\delta_{i0}$, $\delta_{i1}$, $\rho_{i0}^{(mn)}$ and $\rho_{i1}^{(mn)}$ 
depend on the couplings of $J/\psi$, $\psi(2S)$ and the $\Upsilon(mS)$ 
to the channels $\pi\pi$, $K\overline{K}$ and $\eta\eta$. 
The pole term in Eq.~(1) in front of $T_{12}$ is an approximation of possible 
$\phi K$ states, not forbidden by OZI rules. Generally, considering quark diagrams, 
one can see that due to the OZI rules it ought to introduce the pole term in Eq.~(2) 
in front of $T_{22}$. However it is turned out that this pole even a little 
deteriorates the description. Therefore, it is excluded from Eq.~(2).
The numbers in front of square brackets are  coefficients of the vector addition 
of two isospins $I^{(1)}$ and $I^{(2)}$ 
$\bigl(I^{(1)}I^{(2)}I^{(1)}_3I^{(2)}_3\bigl|II_3\bigr)$
where $I$ and $I_3$ are the total isospin and its third component. 
These coefficients are distinct from zero if $I_3=I^{(1)}_3+I^{(2)}_3$. 
The explicit form of relevant coefficient of the vector addition is
\begin{equation}
\Bigl(I^{(1)}I^{(2)}I^{(1)}_3,-I^{(1)}_3\Bigl|00\Bigr)=
(-1)^{I^{(2)}-I^{(1)}_3}\frac{\delta_{I^{(1)}I^{(2)}}}{\sqrt{2I^{(2)}+1}}.
\end{equation}
Then inserting the numerical values of pion and kaon isospins, 
we obtain the corresponding coefficients in Eqs.~(1)-(4).

The amplitudes $T_{ij}$ are expressed through the $S$-matrix elements
\begin{equation}
S_{ij}=\delta_{ij}+2i\sqrt{\rho_1\rho_2}T_{ij}
\end{equation}
where $\rho_i=\sqrt{1-s_i/s}$ and $s_i$ is the reaction threshold. 
The $S$-matrix elements are taken as the products
\begin{equation}\label{S}
S=S^{bgr} S^{res}
\end{equation}
where $S^{res}$ represents the contribution of resonances, 
$S^{bgr}$ is the background part.
The $S^{res}$-matrix elements are parametrized on the uniformization plane 
of the $\pi\pi$-scattering $S$-matrix element by poles and zeros which represent 
resonances. The uniformization plane is obtained by a conformal map of 
the 8-sheeted Riemann surface, on which the three-channel $S$ matrix is determined, 
onto the plane. In the uniformizing variable used~\cite{SBL-prd12}
\begin{equation}
w=\frac{\sqrt{(s-s_2)s_3} + 
\sqrt{(s-s_3)s_2}}{\sqrt{s(s_3-s_2)}}~~~~(s_2=4m_K^2 ~ {\rm and}~ s_3=4m_\eta^2)
\end{equation}
we have neglected the $\pi\pi$-threshold branch point and allowed for 
the $K\overline{K}$- and $\eta\eta$-threshold branch points and left-hand 
branch point at $s=0$ related to the crossed channels. 
Reason of neglecting the $\pi\pi$-threshold branch point consists in following.
With the help of a simple mapping, a function, determined on the 8-sheeted 
Riemann surface, can be uniformized only on torus. 
This is unsatisfactory for our purpose. 
Therefore, we neglect the influence of the lowest ($\pi\pi$) threshold 
branch-point (however, unitarity on the $\pi\pi$-cut is taken into account). 
An approximation like this means the consideration of the nearest to 
the physical region semi-sheets of the Riemann surface of the $S$-matrix. 
In fact, we construct a 4-sheeted model of the initial 8-sheeted Riemann surface
approximating it in accordance with our approach of a consistent account of 
the nearest singularities on all the relevant sheets. 
In practice the disregard of influence of the $\pi\pi$-threshold
branch-point denotes that we do not describe some small region near 
the threshold. This problem was discussed with some details in our earlier works, 
e.g., in~\cite{SBL-prd12}.

Resonance representations on the Riemann surface are obtained using formulas 
from Ref.~\cite{KMS-96}, expressing analytic continuations of the $S$-matrix 
elements to all sheets in terms of those on the physical (I) sheet that have 
only the resonances zeros (beyond the real axis), at least, around the physical 
region. These formulas show how singularities and resonance poles and zeros are 
transferred from the matrix element $S_{11}$ to matrix elements of coupled processes.

The background is introduced to the $S^{bgr}$-matrix elements in a natural way: 
on the threshold of each important channel there appears generally speaking 
a complex phase shift. It is important that we have obtained practically zero 
background of the $\pi\pi$ scattering in the scalar-isoscalar channel. 
This confirms well, first, our assumption $S=S^{bgr}S^{res}$. 
Since in the following combined analysis of the multicannel $\pi\pi$ 
scattering, of the decays $J/\psi\to\phi(\pi\pi,K\overline{K})$, 
$\psi(2S)\to J/\psi\,\pi\pi$, $Y(4260)\to J/\psi~\pi\pi$ and of the two-pion 
transitions of bottomonia state we were not forced to change the resonance-pole 
positions on the Riemann surface of the $S$-matrix and the background parameters, 
we do not give here their values which can be found in other our papers 
(e.g., in Ref.~\cite{SBLKN-prd14}).

Generally, {\it wide multichannel states are most adequately represented by poles}, 
because the poles give the main model-independent effect of resonances and are 
rather stable characteristics for various models, whereas masses and 
total widths are very model-dependent for wide resonances \cite{SKN-epja02}.
The latter and coupling constants of resonances with channels should be calculated 
using the poles on sheets II, IV, and VIII, because only on these sheets the 
analytic continuations have the forms: 
$$\propto 1/S_{11}^{\rm I},~~\propto 1/S_{22}^{\rm I}~~{\rm and}~~\propto 
1/S_{33}^{\rm I},$$
respectively, i.e., the pole positions of resonances are at the same points of 
the complex-energy plane, as the resonance zeros on the physical sheet, 
and are not shifted due to the coupling of channels.

Further, since studying the decays of charmonia and bottomonia, 
we investigated the role of the individual $f_0$ resonances in 
contributing to the shape of the dipion mass distributions in these decays, 
firstly we studied their role in forming the energy dependence of amplitudes 
of reactions $\pi\pi\to\pi\pi,K\overline{K},\eta\eta$.
In this case we switched off only those resonances [$f_0(500)$, $f_0(1370)$,
$f_0(1500)$ and $f_0(1710)$], removal of which can be somehow compensated by
correcting the background (maybe, with elements of the pseudobackground) to have 
the more-or-less acceptable description of the multichannel $\pi\pi$ scattering.
Therefore, below we considered the description of the multichannel $\pi\pi$ 
scattering more for two cases \cite{SBGKLN-pr15_2}:
\begin{itemize}
\item
first, when leaving out a minimal set of the $f_0$ mesons consisting of the 
$f_0(500)$, $f_0(980)$, and $f_0^\prime(1500)$, which is sufficient to achieve 
a description of the processes $\pi\pi\to\pi\pi,K\overline{K},\eta\eta$ with 
a total $\chi^2/\mbox{ndf}\approx1.20$.
\item
Second, from above-indicated three mesons only the $f_0(500)$ can be switched off 
while still obtaining a reasonable description of multichannel $\pi\pi$ scattering 
(though with an appearance of the pseudobackground) with a total 
$\chi^2/\mbox{ndf}\approx1.43$.
\end{itemize}
In Fig.1 we show the obtained description of the processes 
$\pi\pi\!\to\!\pi\pi,K\overline{K},\eta\eta$. 
The solid lines correspond to contribution of all relevant $f_0$-resonances; 
the dotted, of the $f_0(500)$, $f_0(980)$, and $f_0^\prime(1500)$; the dashed, of 
the $f_0(980)$ and $f_0^\prime(1500)$.
\begin{figure}[!thb]
\begin{center}
\includegraphics[width=0.495\textwidth,angle=0]{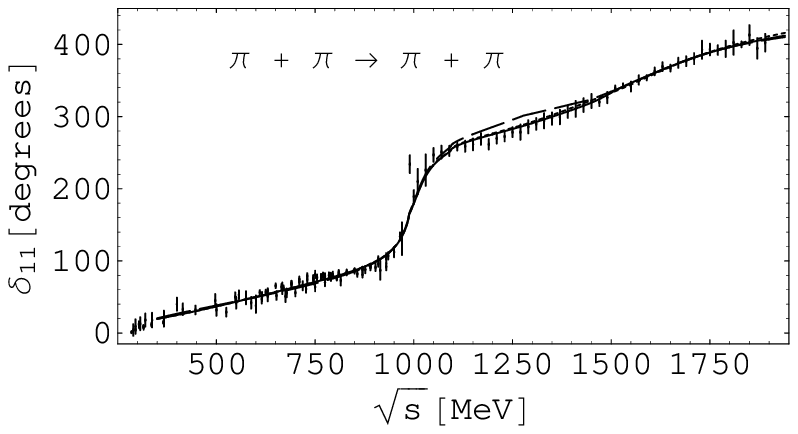}
\includegraphics[width=0.495\textwidth,angle=0]{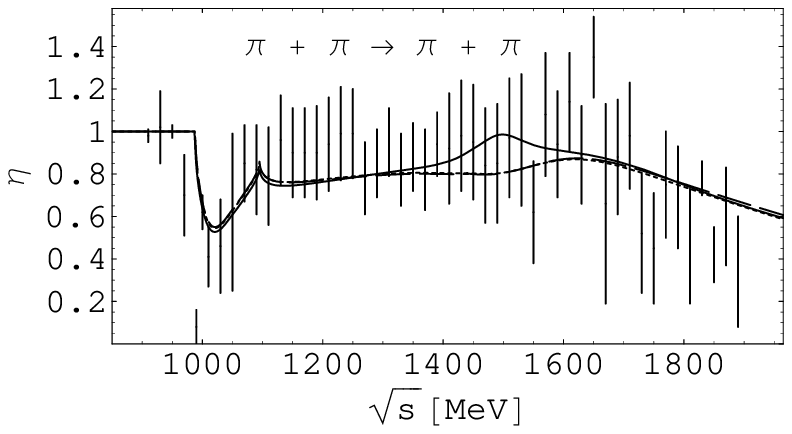}\\
\vspace*{0.12cm}
\includegraphics[width=0.495\textwidth,angle=0]{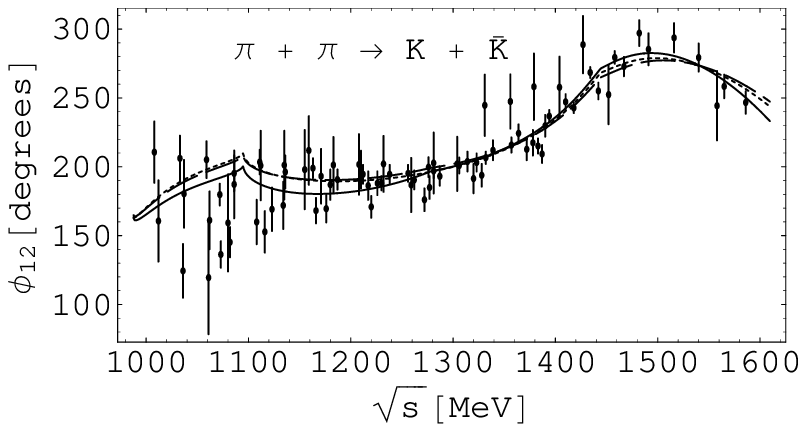}
\includegraphics[width=0.495\textwidth,angle=0]{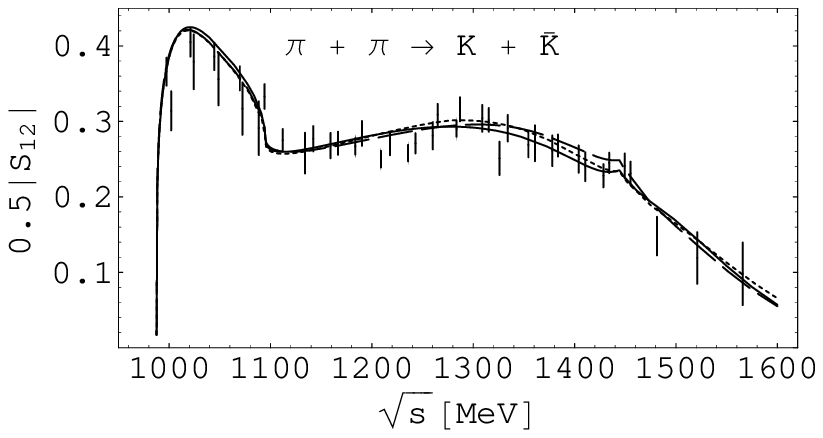}\\
\vspace*{0.12cm}
\includegraphics[width=0.495\textwidth,angle=0]{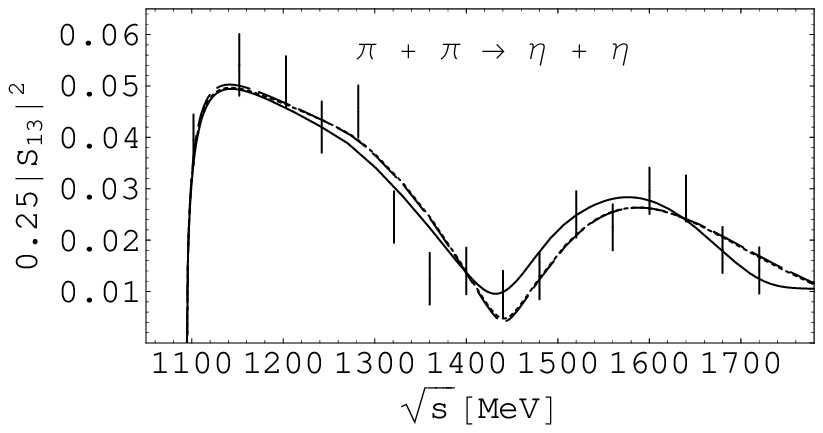}
\vskip -.2cm
\caption{The phase shifts and moduli of the $S$-matrix element in the S-wave 
$\pi\pi$-scattering (upper panel), in $\pi\pi\to K\overline{K}$ (middle panel), 
and the squared modulus of the $\pi\pi\to\eta\eta$ $S$-matrix element (lower figure). 
The solid lines correspond to contribution of all relevant $f_0$-resonances; 
the dotted, of the $f_0(500)$, $f_0(980)$, and $f_0^\prime(1500)$; the dashed, 
of the $f_0(980)$ and $f_0^\prime(1500)$.}
\end{center}\label{fig:fitting}
\end{figure}
One can see that the curves are quite similar in all three cases.

Coming back to the decay analysis, the expression
\begin{equation}
N|F|^{2}\sqrt{(s-s_1)[m_\psi^{2}-(\sqrt{s}-m_\phi)^{2}][m_\psi^2-
(\sqrt{s}+m_\phi)^2]}
\end{equation}
for decays $J/\psi\to\phi(\pi\pi,K\overline{K})$ and the analogues relations 
for $\psi(2S)\to\psi(1S)\pi\pi$~ and $\Upsilon(mS)\to\Upsilon(nS)\pi\pi$ give 
the di-meson mass distributions. $N$ (normalization to experiment) is: 
for $J/\psi\to\phi(\pi\pi,K\overline{K})$ ~1.6663 (Mark~III), 0.5645 (DM~2) 
and 12.1066 (BES~II); for $\psi(2S)\to J/\psi\pi^+\pi^-$ 4.1763 (Mark~II); 
for $\psi(2S)\to J/\psi\pi^0\pi^0$ 3.9825 (Crystal Ball(80)); 
for $\Upsilon(2S)\to \Upsilon(1S)\pi^+\pi^-$ 11.1938 (ARGUS), ~5.6081 (CLEO(94)) 
and 2.9249 (CUSB); for $\Upsilon(2S)\to\Upsilon(1S)\pi^0\pi^0$ 0.6627 (CLEO(07)) 
and 0.2071 (Crystal Ball(85)); 
for $\Upsilon(3S)\to\Upsilon(1S)(\pi^+\pi^-~{\rm and}~\pi^0\pi^0)$ ~57.8466 
and ~13.1958 (CLEO(07)); for
$\Upsilon(3S)\to\Upsilon(2S)(\pi^+\pi^-$ ${\rm and}~\pi^0\pi^0)$ ~6.0706 
and ~4.1026 (CLEO(94)); 
for $\Upsilon(4S)\to\Upsilon(1S)\pi^+\pi^-$ ~13.6322 ({\it BaBar}(06)) 
and ~1.0588 (Belle(07)); 
for $\Upsilon(4S)\to\Upsilon(2S)\pi^+\pi^-$ ~111.418 ({\it BaBar}(06)); 
for $\Upsilon(5S)\to\Upsilon(1S)\pi^+\pi^-$,  
$\Upsilon(5S)\to\Upsilon(2S)\pi^+\pi^-$ and 
$\Upsilon(5S)\to\Upsilon(3S)\pi^+\pi^-$ respectively ~0.6258, 9.1608 
and 20.0786 (Belle(12)); for $\Upsilon(5S)\to\Upsilon(1S)\pi^0\pi^0-$,  
$\Upsilon(5S)\to\Upsilon(2S)\pi^0\pi^0$ and $\Upsilon(5S)\to\Upsilon(3S)\pi^0\pi^0$ 
respectively ~0.2929, 3.0295, and 6.3207 (Belle(13)).

Satisfactory combined description of all considered processes is obtained 
with the total $\chi^2/\mbox{ndf}=842.958/(808 - 122)\approx1.23$; 
for the $\pi\pi$ scattering, $\chi^2/\mbox{ndf}\approx1.14$; 
for $\pi\pi\to K\overline{K}$, $\chi^2/\mbox{ndf}\approx1.65$; 
for $\pi\pi\to\eta\eta$, $\chi^2/\mbox{ndp}\approx0.88$;
for decays $J/\psi\to\phi(\pi^+\pi^-,K\overline{K})$, $\chi^2/\mbox{ndf}\approx1.26$ 
for $\psi(2S)\to J/\psi(\pi^+\pi^-,\pi^0\pi^0)$, $\chi^2/\mbox{ndf}\approx2.74$; 
for $\Upsilon(2S)\to\Upsilon(1S)(\pi^+\pi^-,\pi^0\pi^0)$, $\chi^2/\mbox{ndf}\approx1.07$;
for $\Upsilon(3S)\to\Upsilon(1S)(\pi^+\pi^-,\pi^0\pi^0)$, $\chi^2/\mbox{ndf}\approx1.08$,
for $\Upsilon(3S)\to\Upsilon(2S)(\pi^+\pi^-,\pi^0\pi^0)$, $\chi^2/\mbox{ndf}\approx0.71$,
for $\Upsilon(4S)\to\Upsilon(1S)(\pi^+\pi^-)$, $\chi^2/\mbox{ndf}\approx0.46$, 
for $\Upsilon(4S)\to\Upsilon(2S)(\pi^+\pi^-)$, $\chi^2/\mbox{ndp}\approx0.20$,
for $\Upsilon(5S)\to\Upsilon(1S)(\pi^+\pi^-,\pi^0\pi^0)$, 
$\chi^2/\mbox{ndf}\approx1.39$, 
for $\Upsilon(5S)\to\Upsilon(2S)(\pi^+\pi^-,\pi^0\pi^0)$, 
$\chi^2/\mbox{ndf}\approx1.10$, 
for $\Upsilon(5S)\to\Upsilon(3S)(\pi^+\pi^-,\pi^0\pi^0)$, 
$\chi^2/\mbox{ndf}\approx0.87$.
\vspace*{0.2cm}

The free parameters in Eqs.~(1)-(4), 
depending on the couplings of $J/\psi$, $\psi(2S)$ and the $\Upsilon(mS)$
to the channels $\pi\pi$, $K\overline{K}$ and $\eta\eta$, are found to be
$\alpha_2=0.1729\pm0.011$, $\beta_2=-0.0438\pm0.027$,
$\gamma_{10}= 0.8807\pm 0.023$, $\gamma_{11}= 1.0524\pm 0.016$,
$\gamma_{20}=-2.1591\pm 0.029$, $\gamma_{21}= 0.1419\pm 0.030$,
$\gamma_{30}= 3.0636\pm 0.017$, $\gamma_{31}=-2.6181\pm 0.018$,
$\delta_{10}= 0.5054\pm 0.011$, $\delta_{11}=  9.2480\pm 0.072$,
$\delta_{20}= 6.0865\pm 0.087$, $\delta_{21}=-57.1203\pm 1.890$,
$\delta_{30}=-5.1795\pm 0.032$, $\delta_{31}=  2.6004\pm 0.027$,
$\rho_{10}^{(21)}=0.5117\pm 0.013$, $\rho_{11}^{(21)}= 46.1651\pm 0.656$,
$\rho_{20}^{(21)}=0.6397\pm 0.043$, $\rho_{21}^{(21)}=-21.6599\pm 0.972$,
$\rho_{30}^{(21)}=9.0577\pm 0.054$, $\rho_{31}^{(21)}= -6.3357\pm 0.034$,
$\rho_{10}^{(31)}=0.9728\pm 0.026$, $\rho_{11}^{(31)}=-2.4287\pm 0.023$,
$\rho_{20}^{(31)}=0.9164\pm 0.041$, $\rho_{21}^{(31)}= 0.5073\pm 0.017$,
$\rho_{30}^{(31)}=0.2223\pm 0.017$, $\rho_{31}^{(31)}=-0.7330\pm 0.021$,
$\rho_{10}^{(32)}= 0.0274\pm 0.010$, $\rho_{11}^{(32)}= 56.4752\pm 1.770$,
$\rho_{20}^{(32)}= 1.6553\pm 0.027$, $\rho_{21}^{(32)}=-50.5964\pm 2.600$,
$\rho_{30}^{(32)}=16.1394\pm 0.870$, $\rho_{31}^{(32)}=-55.0251\pm 2.130$,
$\rho_{10}^{(41)}= 0.4889\pm 0.012$, $\rho_{11}^{(41)}=-2.4299\pm 0.057$,
$\rho_{20}^{(41)}=-0.8203\pm 0.041$, $\rho_{21}^{(41)}= 0.0583\pm 0.012$,
$\rho_{30}^{(41)}=-0.0791\pm 0.011$, $\rho_{31}^{(41)}= 0.0542\pm 0.003$,
$\rho_{10}^{(42)}=2.5852\pm 0.066$, $\rho_{11}^{(42)}=-8.7188\pm 0.177$,
$\rho_{20}^{(42)}=1.7985\pm 0.059$, $\rho_{21}^{(42)}=-9.7334\pm 0.790$,
$\rho_{30}^{(42)}=0.6851\pm 0.019$, $\rho_{31}^{(42)}= 0.9233\pm 0.035$,
$\rho_{10}^{(51)}=-1.1574\pm 0.057$, $\rho_{11}^{(51)}= 5.1800\pm 0.221$,
$\rho_{20}^{(51)}= 3.7654\pm 0.033$, $\rho_{21}^{(51)}=-4.7934\pm 0.834$,
$\rho_{30}^{(51)}=-3.0899\pm 0.054$, $\rho_{31}^{(51)}= 1.9762\pm 0.065$,
$\rho_{10}^{(52)}= 1.2657\pm 0.063$, $\rho_{11}^{(52)}= 1.4487\pm 0.071$,
$\rho_{20}^{(52)}=-1.3707\pm 0.057$, $\rho_{21}^{(52)}= 2.2858\pm 0.770$,
$\rho_{30}^{(52)}=-5.6127\pm 0.041$, $\rho_{31}^{(52)}=10.6278\pm 1.120$,
$\rho_{10}^{(53)}= 1.0362\pm 0.016$, $\rho_{11}^{(53)}=1.8643\pm 0.047$,
$\rho_{20}^{(53)}= 0.6141\pm 0.023$, $\rho_{21}^{(53)}=0.1688\pm 0.063$,
$\rho_{30}^{(53)}=-0.5437\pm 0.019$, $\rho_{31}^{(53)}=0.3827\pm 0.071$.

In Figs.~2-7 we shown our fits to the decay data. In all next figures  
the solid lines, as above, correspond to contribution
of all relevant $f_0$-resonances; the dotted, of the $f_0(500)$, $f_0(980)$,
and $f_0^\prime(1500)$; the dashed, of the $f_0(980)$ and $f_0^\prime(1500)$.
\begin{figure}[!thb]
\begin{center}
\includegraphics[width=0.54\textwidth,angle=0]{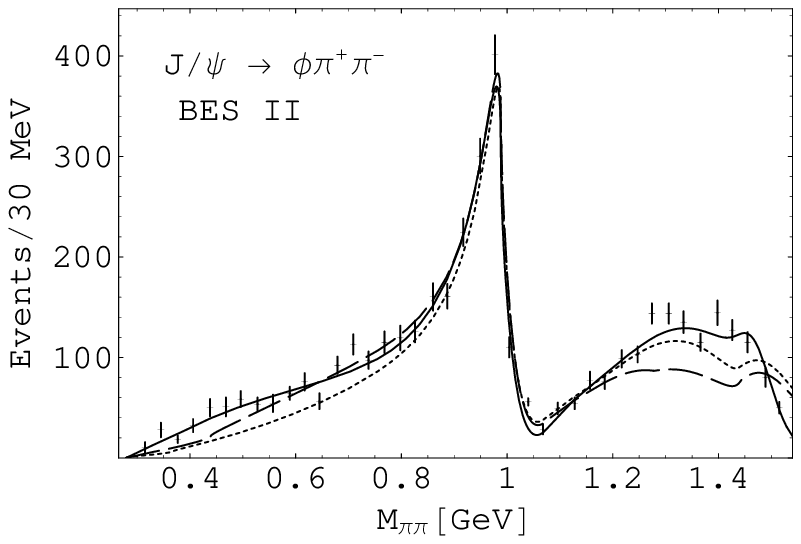}\\
\vspace*{0.12cm}
\includegraphics[width=0.47\textwidth,angle=0]{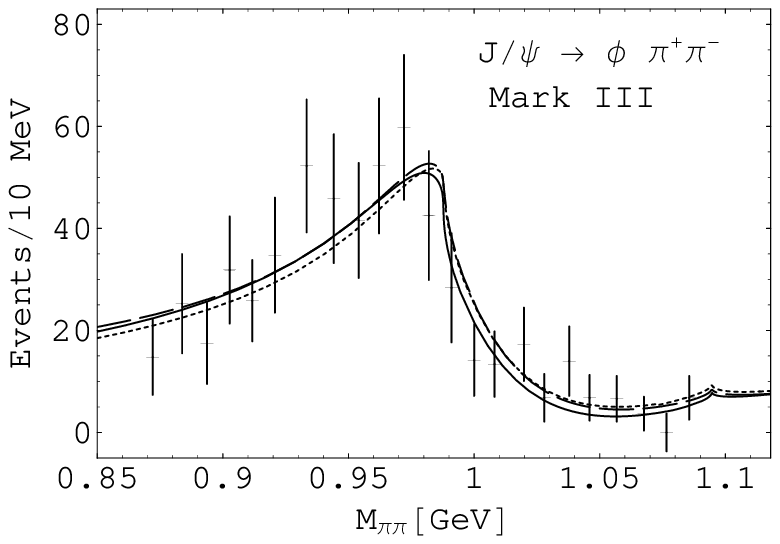}
\includegraphics[width=0.47\textwidth,angle=0]{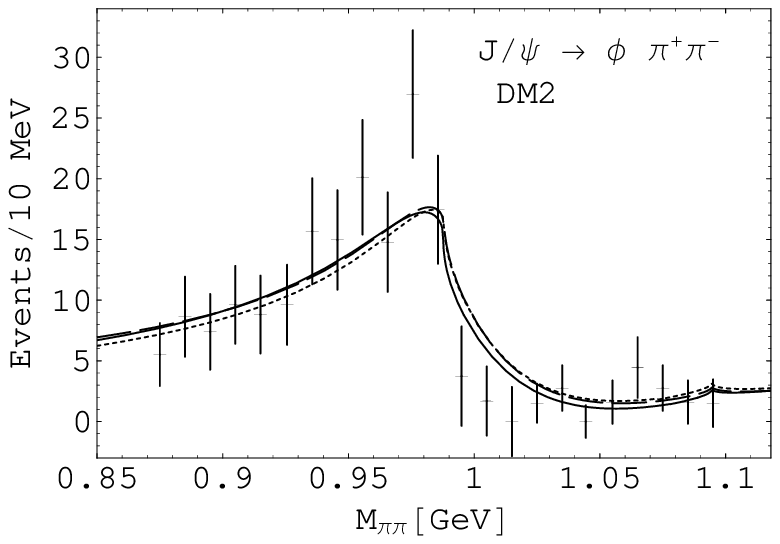}\\
\vspace*{0.12cm}
\includegraphics[width=0.47\textwidth,angle=0]{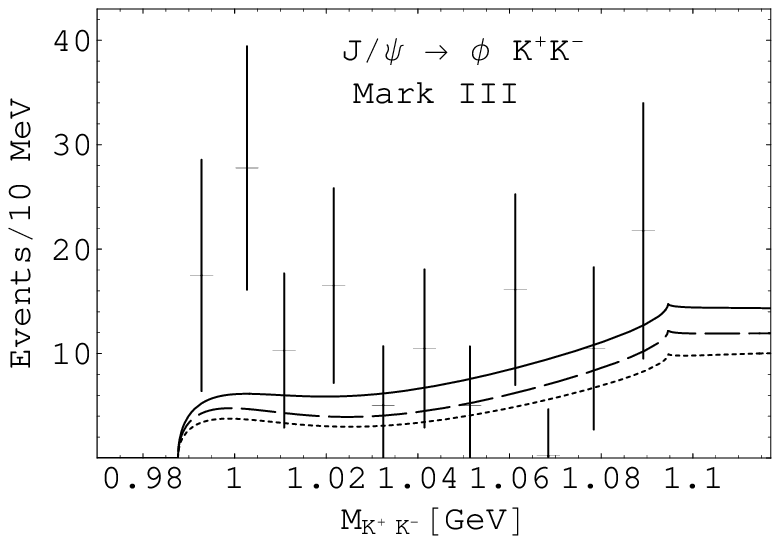}
\includegraphics[width=0.47\textwidth,angle=0]{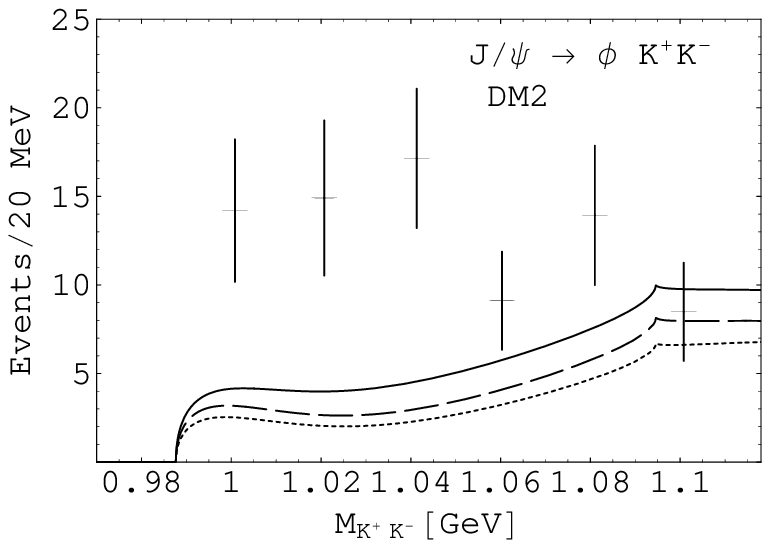}
\vspace*{-0.1cm}\caption{The decays $J/\psi\to\phi(\pi^+\pi^-,K^+K^-)$. 
The solid, dotted and dashed lines as explained in Fig.~1.
}
\end{center}\label{fig:Jphi}
\end{figure}
\begin{figure}[!ht]
\begin{center}
%\vspace*{0.3cm}
\includegraphics[width=0.48\textwidth]{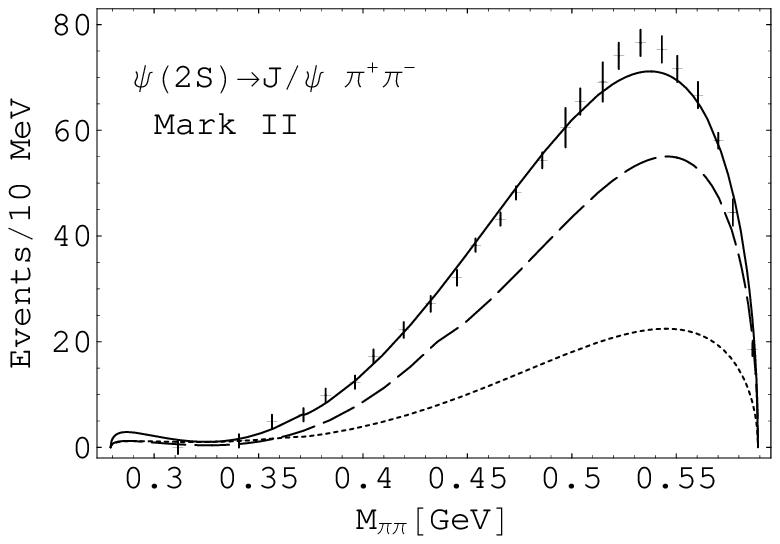}
\includegraphics[width=0.48\textwidth]{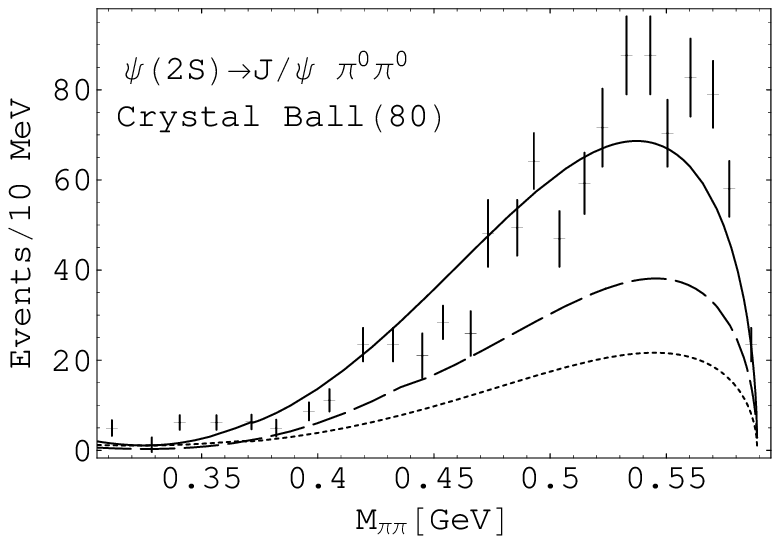}
\vspace*{-0.1cm}
\caption{The decays $\psi(2S)\to J/\psi(\pi^+\pi^-,\pi^0\pi^0)$. 
The solid, dotted and dashed lines as explained in Fig.~1.}
\end{center}\label{fig:BESII}
\end{figure}
\begin{figure}[!ht]
\begin{center}
\includegraphics[width=0.47\textwidth]{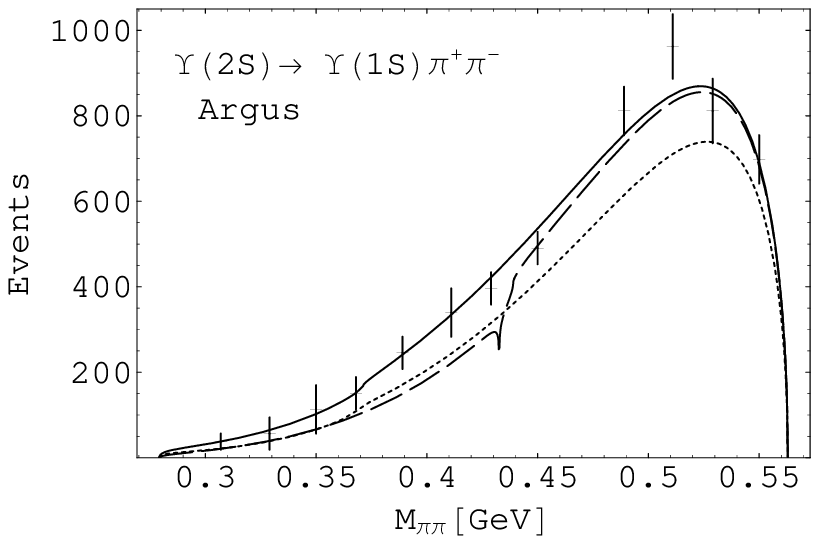}
\includegraphics[width=0.47\textwidth]{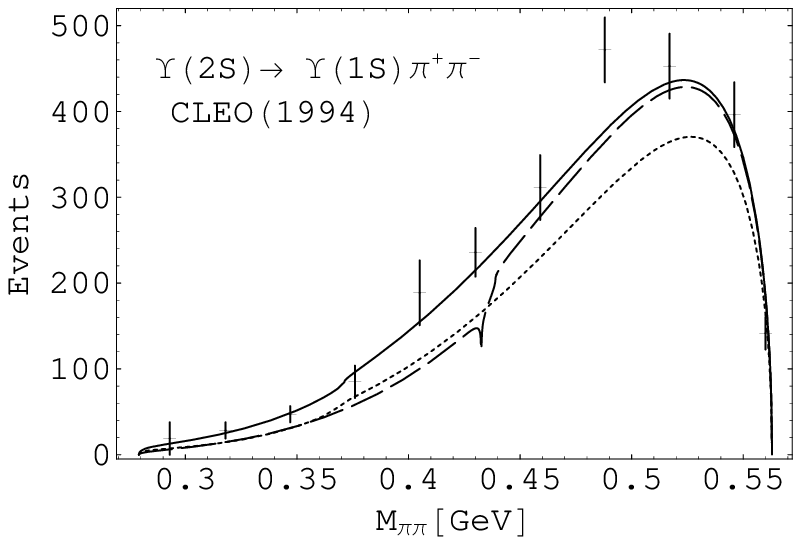}\\
\vspace*{0.12cm}
\includegraphics[width=0.47\textwidth]{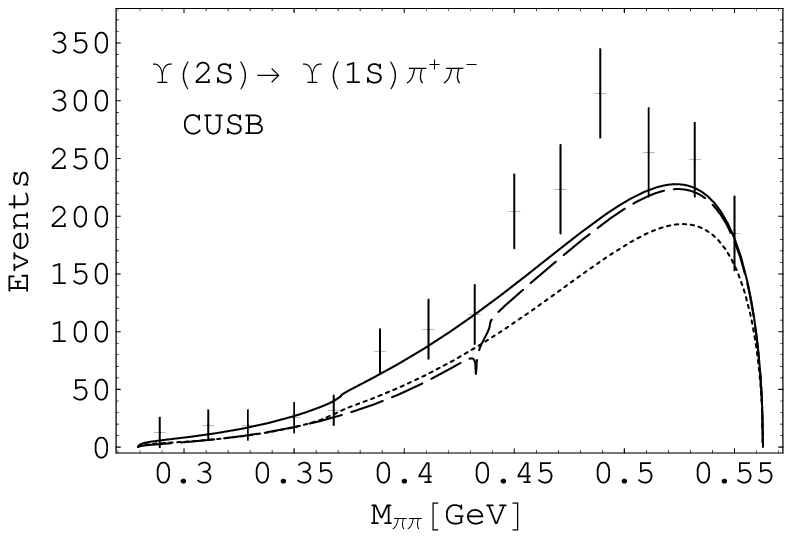}
\includegraphics[width=0.47\textwidth]{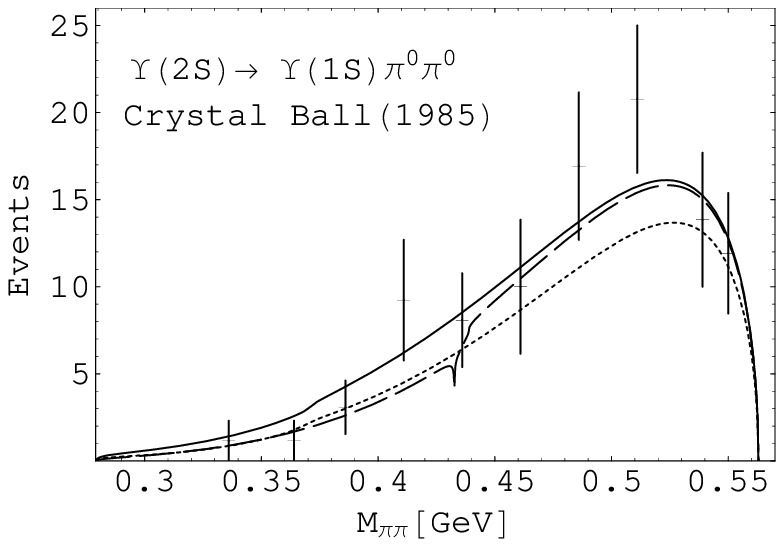}\\
\vspace*{0.12cm}
\includegraphics[width=0.47\textwidth]{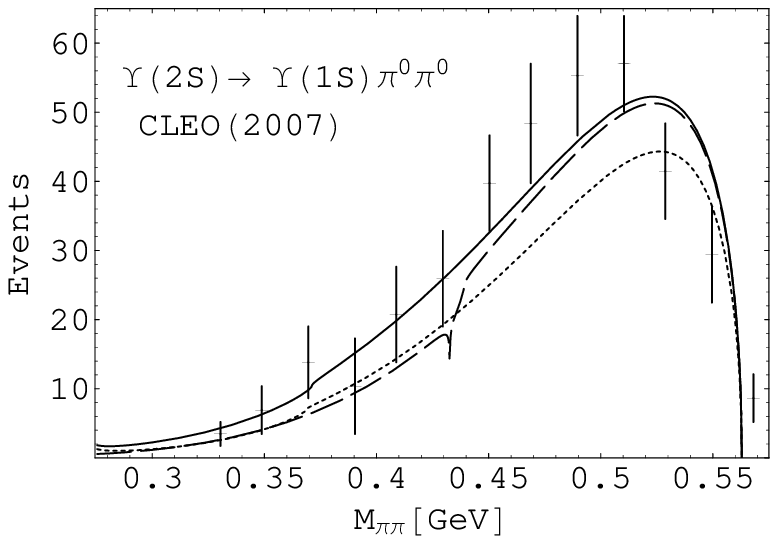}
\vspace*{-0.1cm}\caption{The decays 
$\Upsilon(2S)\to\Upsilon(1S)(\pi^+\pi^-,\pi^0\pi^0)$. 
The solid, dotted and dashed lines as explained in Fig.~1.
}
\end{center}\label{fig:Ups21}
\end{figure}
\begin{figure}[!ht]
\begin{center}
\includegraphics[width=0.47\textwidth]{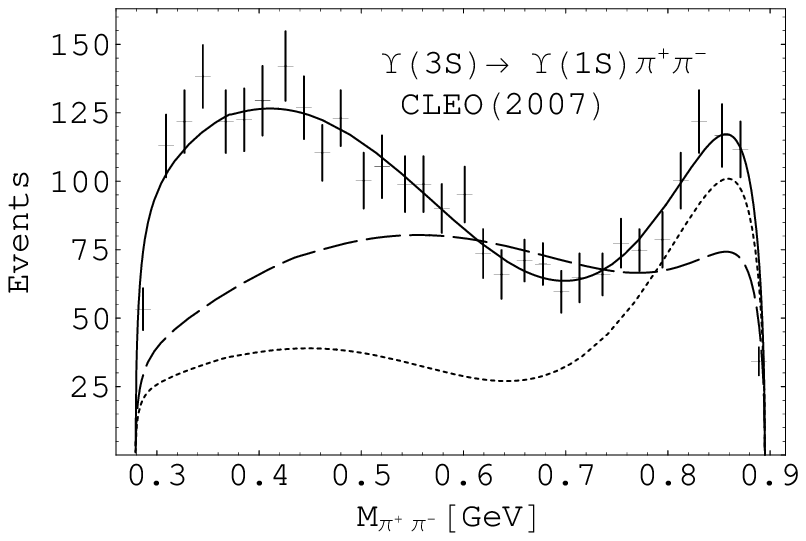}
\includegraphics[width=0.47\textwidth]{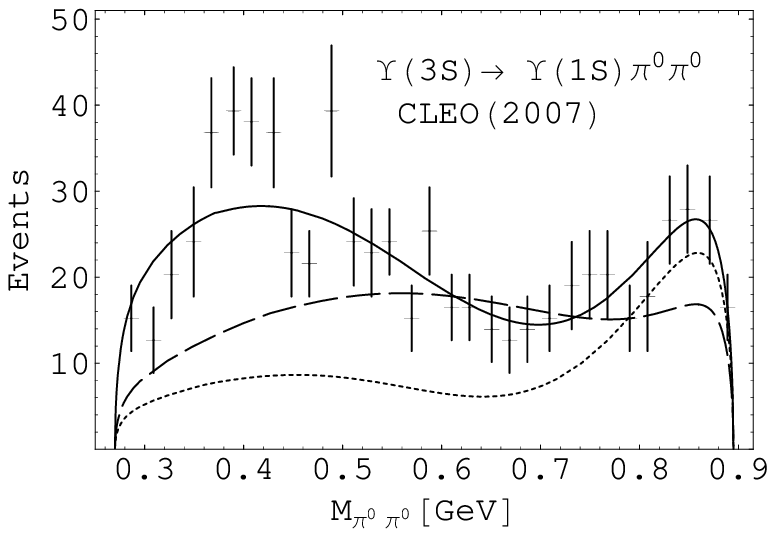}\\
\vspace*{0.12cm}
\includegraphics[width=0.47\textwidth]{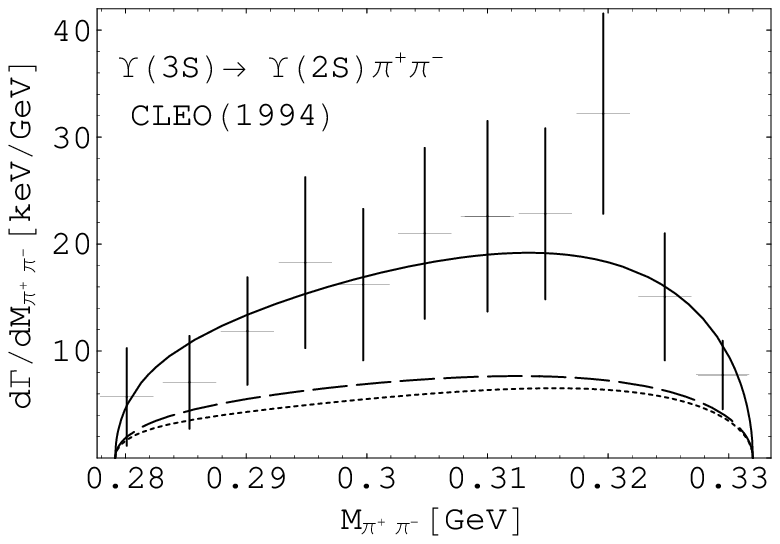}
\includegraphics[width=0.47\textwidth]{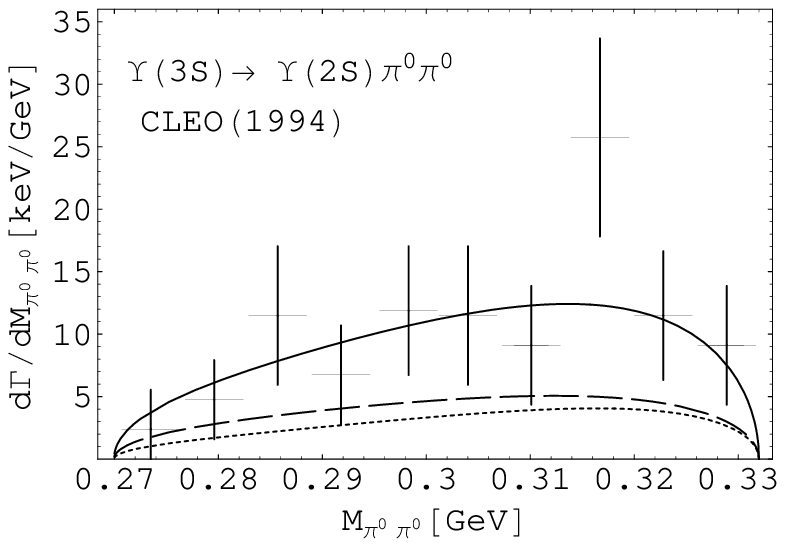}
\vspace*{-0.1cm}\caption{The decays 
$\Upsilon(3S)\to\Upsilon(1S)(\pi^+\pi^-,\pi^0\pi^0)$ (upper panel) and 
$\Upsilon(3S)\to\Upsilon(2S)(\pi^+\pi^-,\pi^0\pi^0)$ (lower panel). 
The solid, dotted and dashed lines as explained in Fig.~1.
}
\end{center}\label{fig:Ups31}
\end{figure}
%\vspace*{-0.15cm}
\begin{figure}[!ht]
\begin{center}
\includegraphics[width=0.45\textwidth]{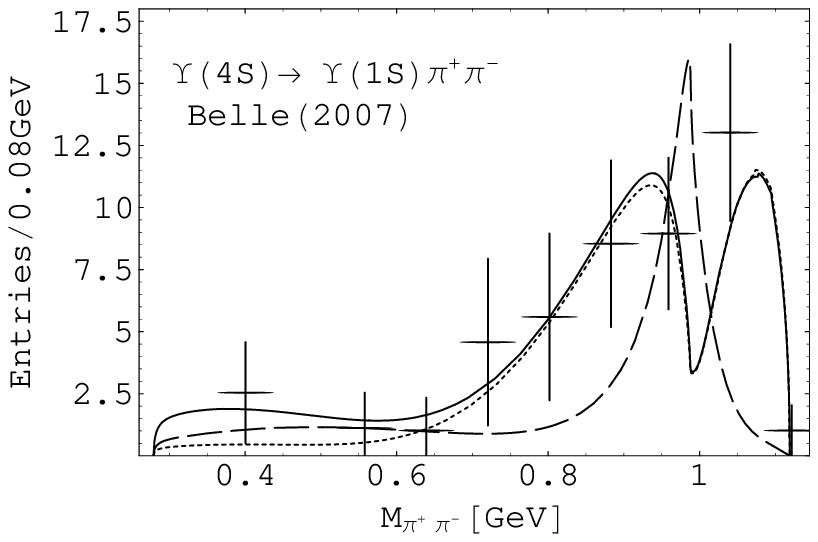}
\includegraphics[width=0.45\textwidth]{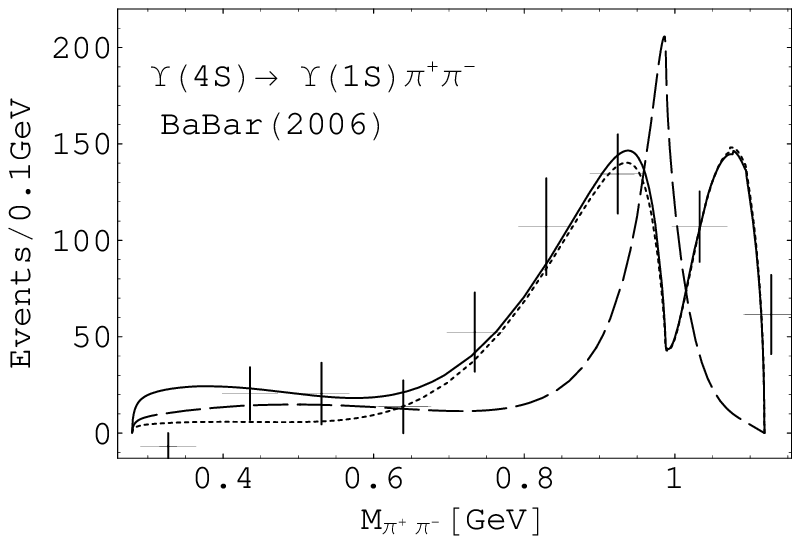}\\
\vspace*{0.14cm}
\includegraphics[width=0.45\textwidth]{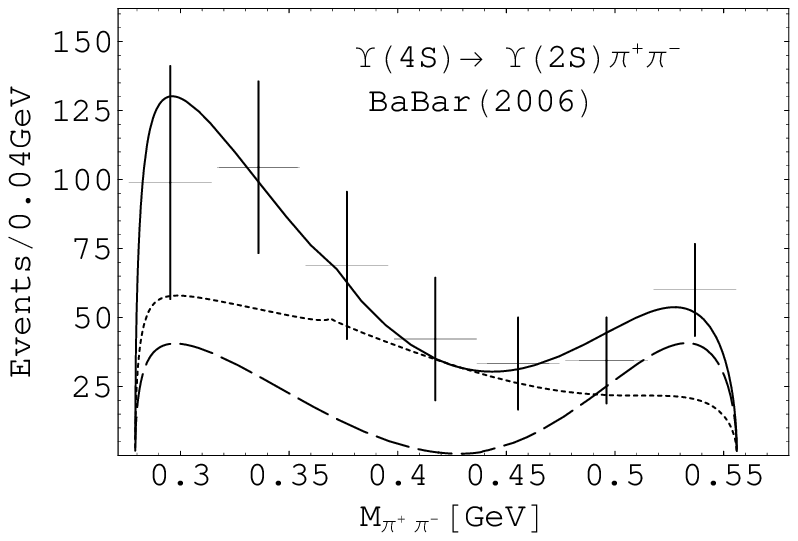}
\vspace*{-0.1cm}\caption{The decays $\Upsilon(4S)\to\Upsilon(1S,2S)\pi^+\pi^-$. 
The solid, dotted and dashed lines as explained in Fig.~1.}
\end{center}\label{fig:Ups4n}
\end{figure}
%
%\vspace*{-0.15cm}
\begin{figure}[!ht]
\begin{center}
\includegraphics[width=0.45\textwidth]{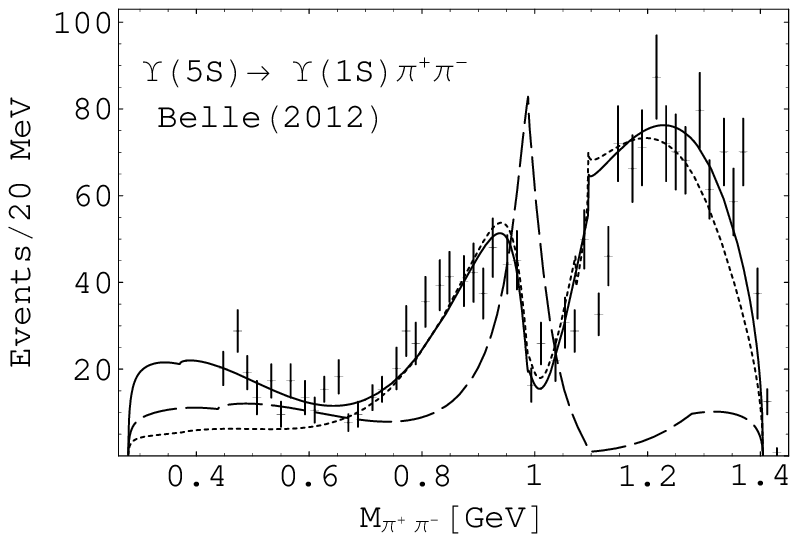}
\includegraphics[width=0.45\textwidth]{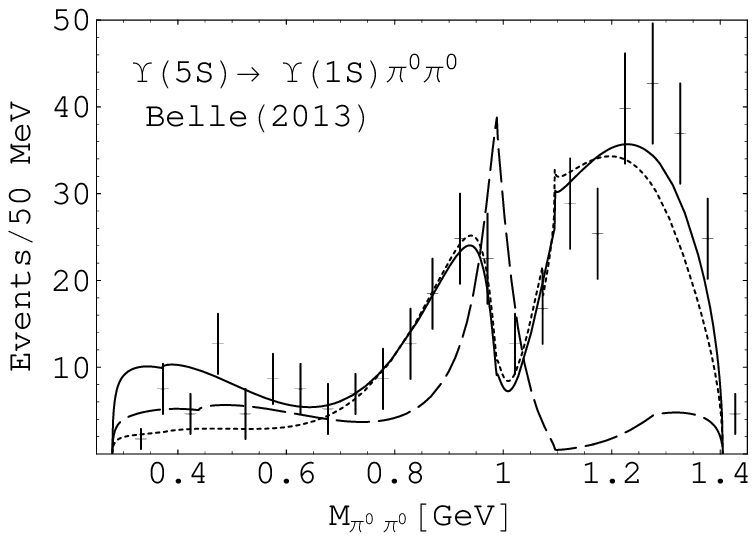}\\
\vspace*{0.12cm}
\includegraphics[width=0.45\textwidth]{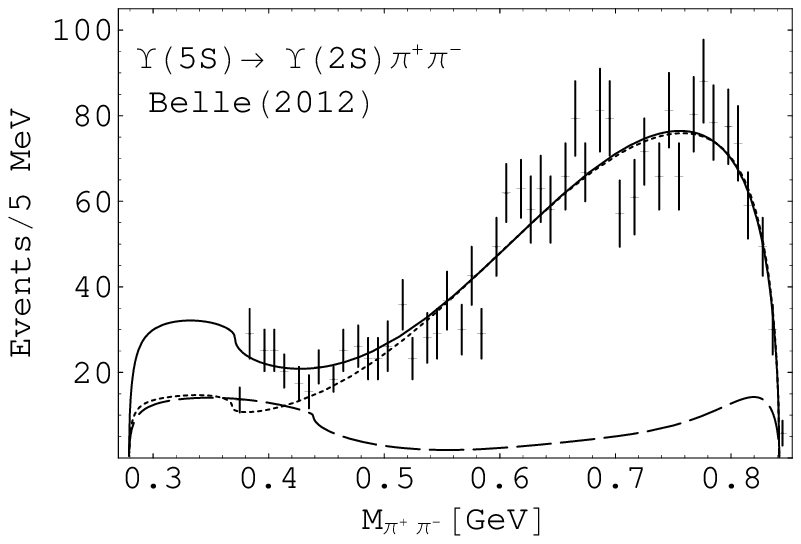}
\includegraphics[width=0.45\textwidth]{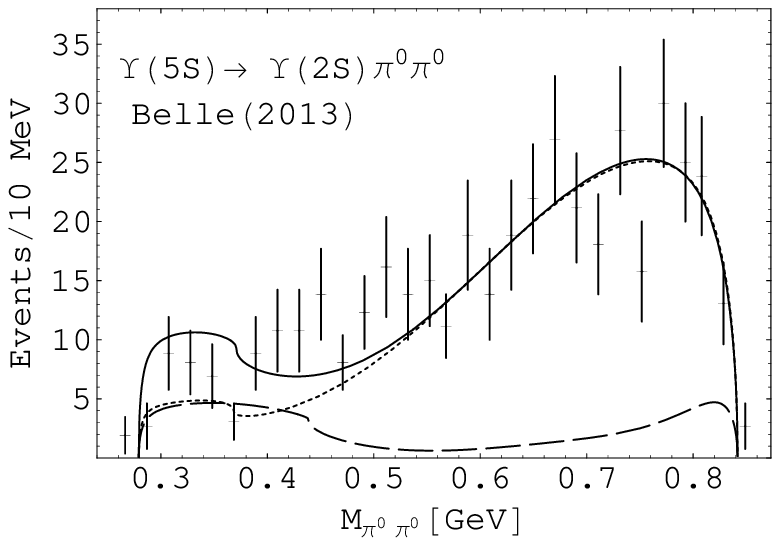}\\
\vspace*{0.12cm}
\includegraphics[width=0.45\textwidth]{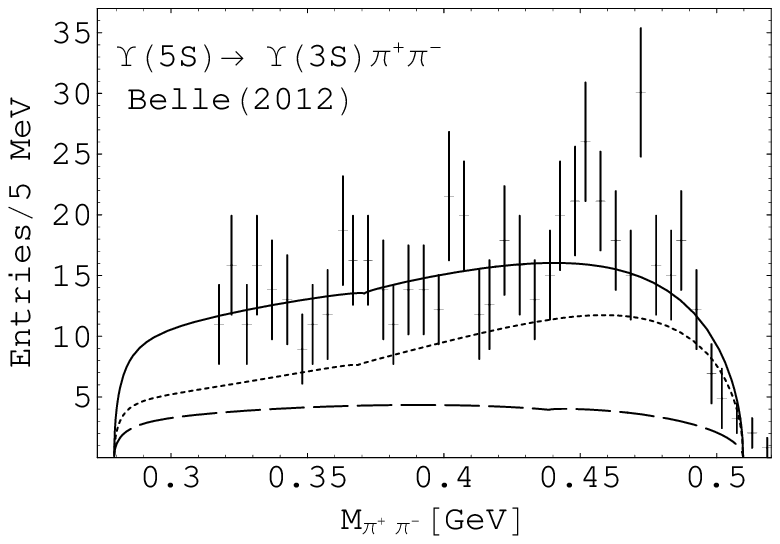}
\includegraphics[width=0.45\textwidth]{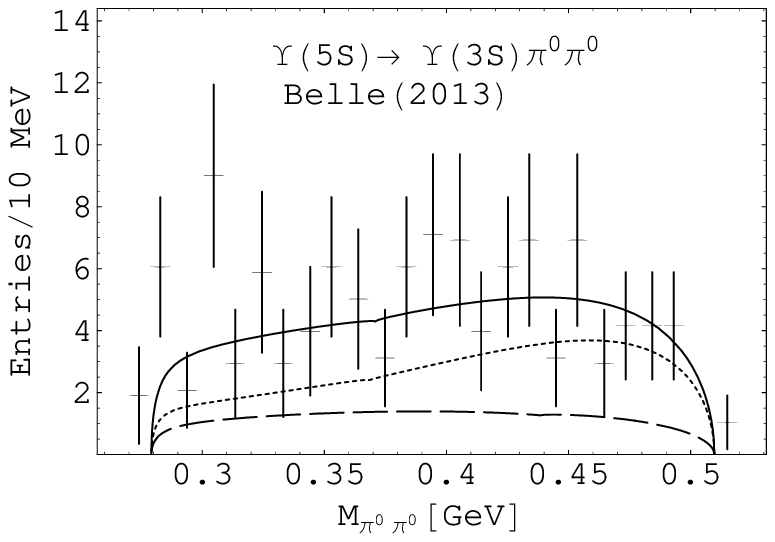}
\vspace*{-0.1cm}\caption{The decays and 
$\Upsilon(5S)\to\Upsilon(ns)(\pi^+ \pi^-,\pi^0\pi^0)$ ($n=1,2,3$). 
The solid, dotted and dashed lines as explained in Fig.~1.}
\end{center}\label{fig:Ups5n}
\end{figure}

Taking into account success in describing the multichannel $\pi\pi$ scattering 
and the above-shown decays of charmonia and bottomonia, 
it is worth to show obtained predictions for amplitudes of the $\eta\eta$ and 
$K\overline{K}$ scattering and of the transitions $\eta\eta\to\pi\pi$ and 
$\eta\eta\to K\overline{K}$ which are used in our calculations and almost or 
entirely unknown from experiment (Fig.~8). In the Appendix we show formulas for 
the phase shifts and moduli of indicated amplitudes which were used also at 
calculating decays of charmonia and bottomonia.
\begin{figure}[!thb]
\begin{center}
\includegraphics[width=0.495\textwidth,angle=0]{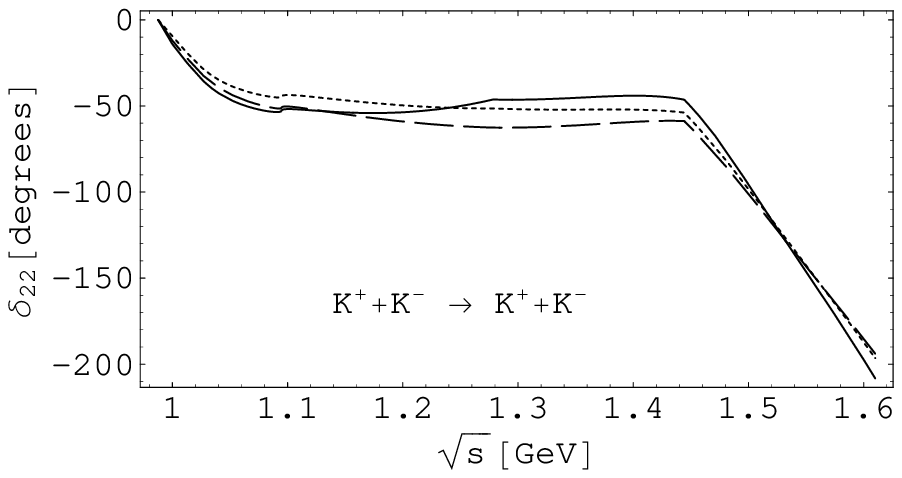}
\includegraphics[width=0.495\textwidth,angle=0]{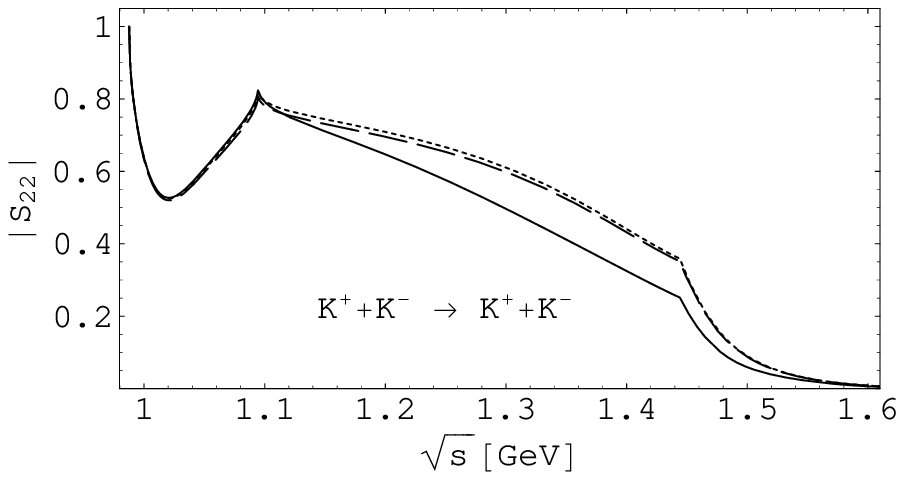}\\
\vspace*{0.12cm}
\includegraphics[width=0.495\textwidth,angle=0]{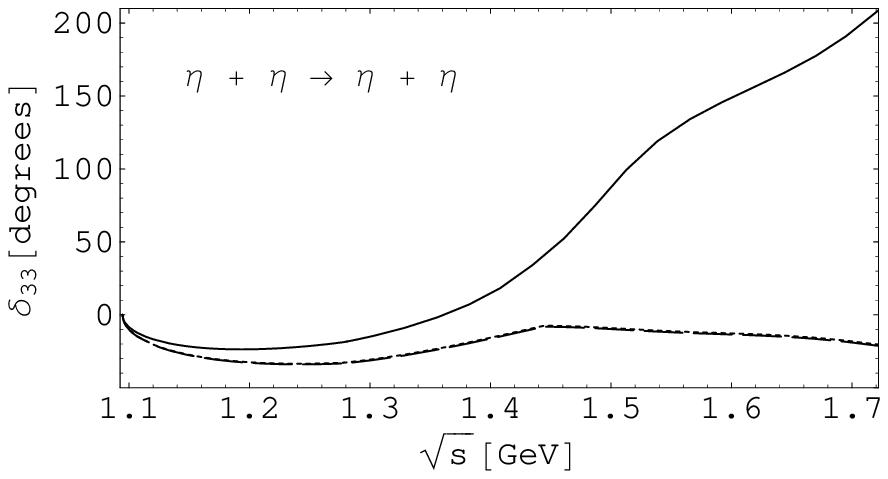}
\includegraphics[width=0.495\textwidth,angle=0]{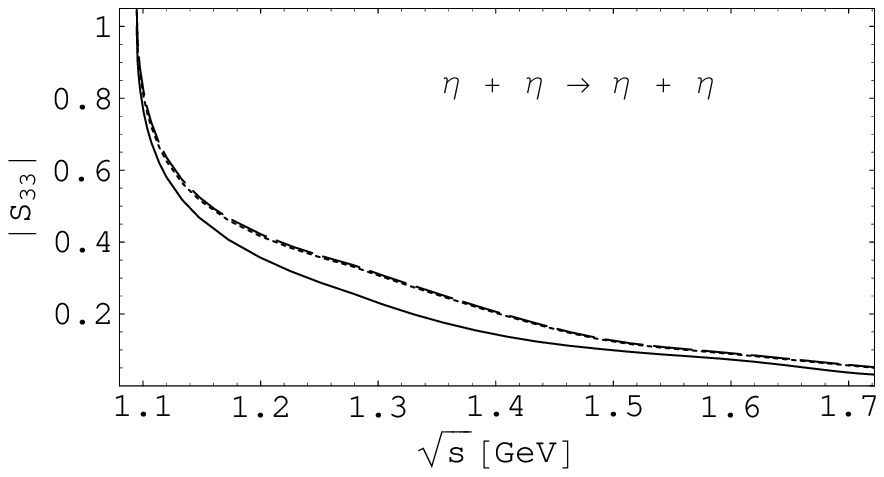}\\
\vspace*{0.12cm}
\includegraphics[width=0.495\textwidth,angle=0]{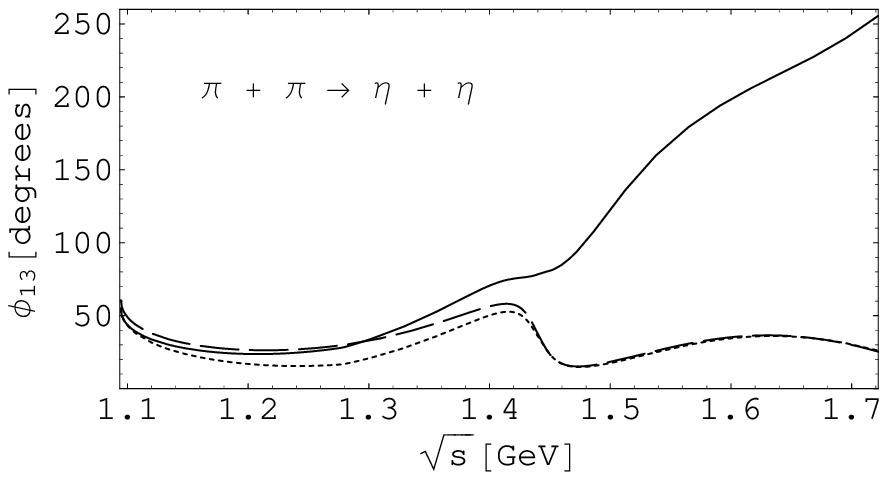}
\includegraphics[width=0.495\textwidth,angle=0]{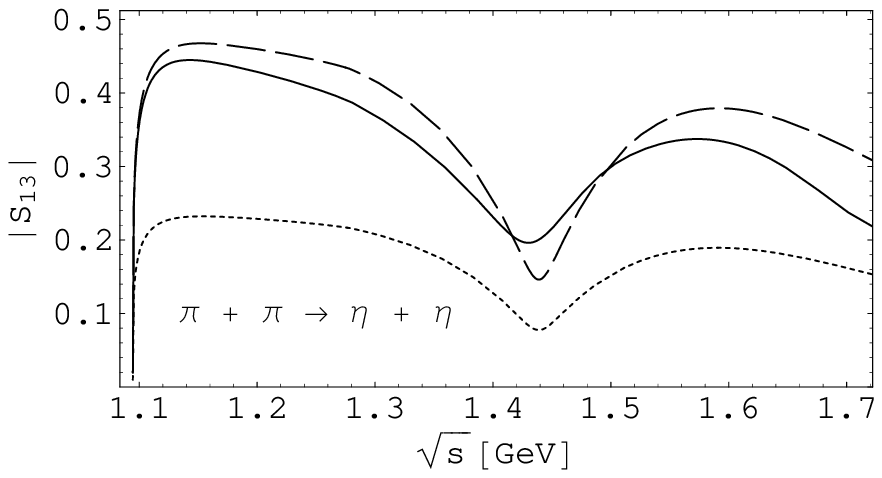}\\
\vspace*{0.12cm}
\includegraphics[width=0.495\textwidth,angle=0]{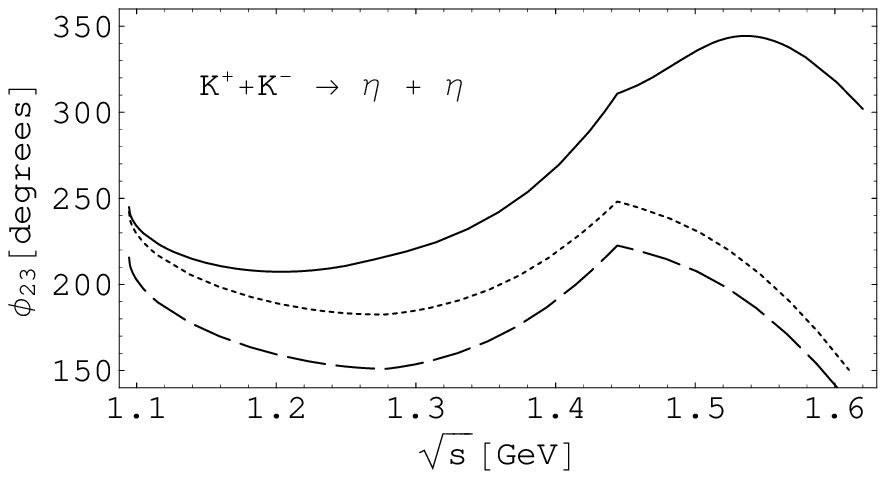}
\includegraphics[width=0.495\textwidth,angle=0]{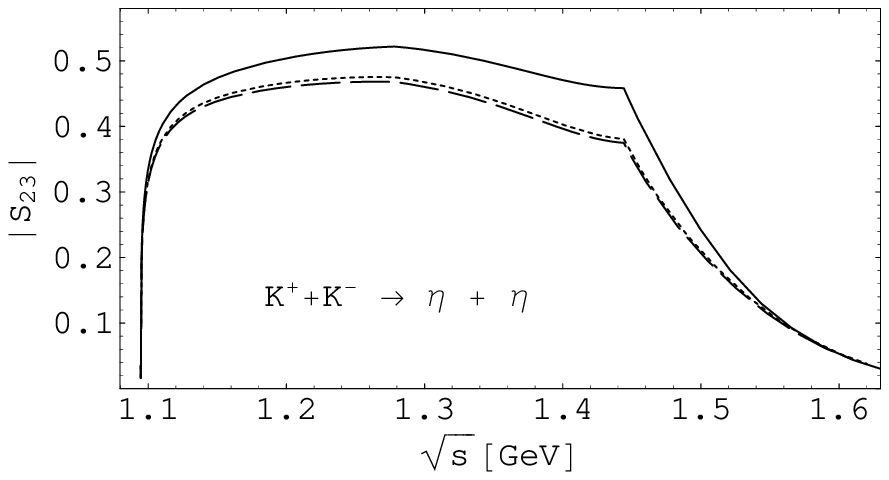}
\vskip -.2cm
\caption{The phase shifts and moduli of the $S$-matrix element 
in the S-wave $K\overline{K}$ and $\eta\eta$ scattering (two upper panels), 
in $\pi\pi\to\eta\eta$ (third panel), 
and in $K\overline{K}\to\eta\eta$ (lower panel). 
The solid, dotted and dashed lines as explained in Fig.~1.}
\end{center}\label{fig:predictions}
\end{figure}

Finally, we have applied our method for describing the data on the decay of 
charmonium $X(4260)$ (sometimes is indicated as $Y(4260)$) to 
$J/\psi~\pi^+\pi^-$ \cite{Belle13}. There was used the formula analogous Eq.~(3), 
and the obtained description is quite satisfactory: 
$\chi^2/\mbox{ndf}\approx1.23$ and fitting to the data shown on Fig.~9. 
The general combined description of all considered processes is obtained 
with the total $\chi^2/\mbox{ndf}=764.417/(739 - 119)\approx1.31$.
\begin{figure}[!ht]
\begin{center}
%\vspace*{0.3cm}
\includegraphics[width=0.48\textwidth]{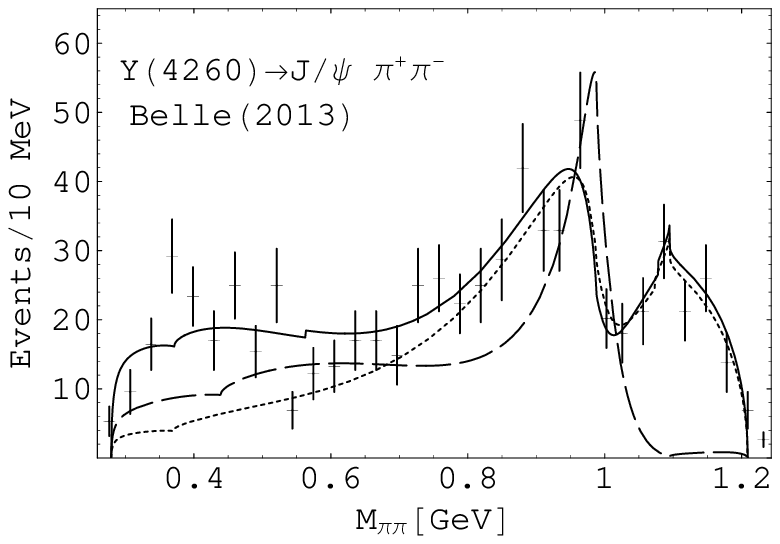}
\includegraphics[width=0.48\textwidth]{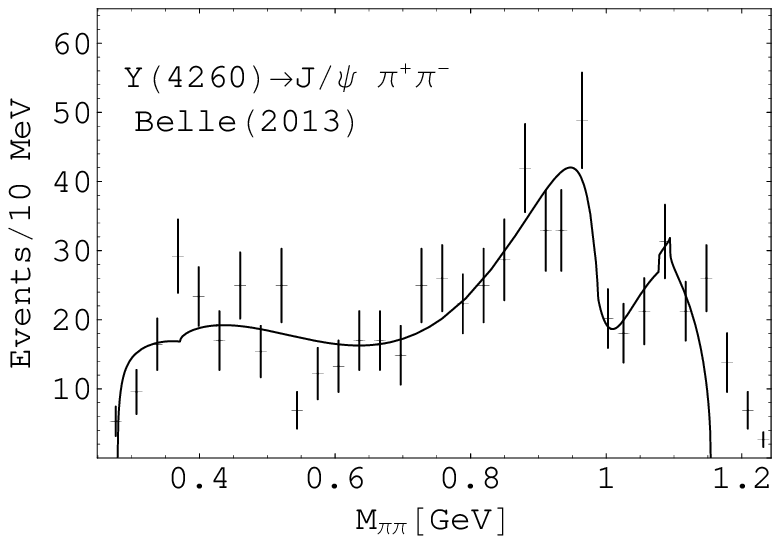}
%\vspace*{-0.2cm}
\caption{The decay $Y(4260)\to J/\psi~\pi^+\pi^-$. 
The solid, dotted and dashed lines as explained in Fig.~1. 
The data are taken from Ref.~\cite{Belle13}.}
\end{center}\label{fig:Y(4260)decay}
\end{figure}
In the PDG tables~\cite{PDG-16} for the state $X(4260)$ 
one indicates the quantum numbers $I^G(J^{PC})=?^?(1^{--})$ 
and the mass $m=4251\pm9$~MeV. 
However, this analysis shows that the data~\cite{Belle13} correspond 
to the decay of charmonium with the mass 4.3102~GeV (left-hand picture), 
not with 4.251~GeV (right-hand one on Fig.~9). Furthermore, since we have 
shown that the basic forms of the dipion mass spectra of charmonia and 
bottomonia pion-pion transitions are explained by the unified mechanism, 
one can think that characteristic pictures of the mass spectra of analogous 
charmonia and bottomonia transitions are similar, of course, with taking into 
account distortions due to the phase space volume. 
Obviously, for justification of this assumption there is important 
the spectator role of vector meson in the final state. 
Furthermore, to some extent this assumption is supported by comparison of 
the corresponding experimental data: cf. Figs.~3 and 4 for 
$\psi(2S)\to J/\psi(1S)(\pi^+\pi^-,\pi^0\pi^0)$ and 
$\Upsilon(2S)\to\Upsilon(1S)(\pi^+\pi^-,\pi^0\pi^0)$, respectively. 
Further one can see that the basic forms of dipion mass spectra of the decay 
$\Upsilon(4S)\to\Upsilon(1S)\pi^+\pi^-$ (Fig.~6, two left-hand pictures from above) 
and of the charmonium $X(4260)$ (Fig.~9) are similar. 
This can be some indication that the $X(4260)$ is a third radial excitation, 
i.e. the $4S$ state with the mass 4.3102~GeV.
The obtained parameters in equation of type (3), which depend on the couplings of 
$X(2S)(4310)$ to the channels $\pi\pi$, $K\overline{K}$ and $\eta\eta$, are
$\delta_{10}^{(41)}=0.0062$, $\delta_{11}^{(41)}=3.6752$, 
$\delta_{10}^{(41)}=4.1488$, $\delta_{11}^{(41)}=-2.7138$, 
$\delta_{10}^{(41)}=-6.2914$, $\delta_{11}^{(41)}=5.5438$.
 $N$ (normalization to experiment) in Eq.~(9) is 0.3567.

\section{Conclusions and Discussion}

The combined analysis was performed for data on isoscalar S-wave processes
$\pi\pi\to\pi\pi,K\overline{K},\eta\eta$ and on the decays of the charmonia ---
$J/\psi\to\phi(\pi\pi,K\overline{K})$, $\psi(2S)\to J/\psi\,\pi\pi$, 
$Y(4260)\to J/\psi~\pi^+\pi^-$ --- and of the bottomonia ---
$\Upsilon(mS)\to\Upsilon(nS)\pi\pi$ ($m>n$, $m=2,3,4,5,$ $n=1,2,3$) 
from the ARGUS, Crystal Ball, CLEO, CUSB, DM2, Mark~II, Mark~III, BES~II, 
{\it BABAR}, and Belle Collaborations.

It is shown that the dipion and $K\overline{K}$ mass spectra in 
the above-indicated decays of charmonia and the dipion mass spectra 
of bottomonia are explained by the unified mechanism which is based on our previous
conclusions on wide resonances~\cite{SBLKN-jpgnpp14,SBLKN-prd14} and is related
to contributions of the $\pi\pi$, $K\overline{K}$ and $\eta\eta$ coupled channels 
including their interference. It is shown that in the final states of these decays 
(except $\pi\pi$ scattering) the contribution of coupled processes, e.g., 
$K\overline{K},\eta\eta\to\pi\pi$, is important even if these processes are 
energetically forbidden.

When analyzing the decay $Y(4260)\to J/\psi~\pi^+\pi^-$, it is obtained some 
indication that the charmonium $X(4260)$, the dipion spectrum of 
$J/\psi~\pi^+\pi^-$ decay of which is published in Ref.~\cite{Belle13}, 
is a third radial excitation, i.e. the $4S$ state with the mass 4.3102~GeV.

The allowance for the effect of the $\eta\eta$ channel in the considered decays both 
kinematically (i.e. via the uniformizing variable) and also by adding the 
$\pi\pi\to\eta\eta$ amplitude in the formulas for the decays permits us to 
eliminate unphysical (i.e. those related with no channel thresholds) 
nonregularities 
in some $\pi\pi$ distributions, being present without this extension of the 
description~\cite{SBGKLN-pr15_2}, obtaining a reasonable and satisfactory description of all considered $\pi\pi$ and $K\overline{K}$ spectra in the two-pion and $K\overline{K}$ 
transitions of charmonia and in the two-pion transitions of bottomonia.

It was also very useful to consider the role of individual $f_0$ resonances in 
contributions to the dipion mass distributions in the indicated decays. For example, 
it is seen that the sharp dips about 1~GeV in the 
$\Upsilon(4S,5S)\to\Upsilon(1S)\pi^+\pi^-$ decays are related with the $f_0(500)$ 
contribution to the interfering amplitudes of $\pi\pi$ scattering and 
$K\overline{K},\eta\eta\to\pi\pi$ processes. Namely the consideration of this role 
of the $f_0(500)$ allows us to make conclusion on the existence of the sharp dip 
about 1~GeV in the dipion mass spectrum of the 
$\Upsilon(4S)\to\Upsilon(1S)\pi^+\pi^-$ decay where, 
unlike $\Upsilon(5S)\to\Upsilon(1S)\pi^+\pi^-$, 
the scarce data do not permit to do that conclusion yet.

Also, a manifestation of the $f_0(1370)$ is turned out to be interesting and 
unexpected. First, in the satisfactory description of the $\pi\pi$ spectrum of 
decay $J/\psi\to\phi\pi\pi$, the second large peak in the 1.4-GeV region can be 
naively imagined as related to the contribution of the $f_0(1370)$.
We have shown that this is not right -- the constructive interference between 
the contributions of the $\eta\eta$ and $\pi\pi$ and $K\overline{K}$ channels 
plays the main role in formation of the 1.4-GeV peak. This is quite in agreement 
with our earlier conclusion that the $f_0(1370)$ has a dominant $s{\bar s}$ 
component~\cite{SBLKN-jpgnpp14}.

On the other hand, it turned out that the $f_0(1370)$ contributes considerably 
in the near-$\pi\pi$-threshold region of many dipion mass distributions, 
especially making the threshold bell-shaped form of the dipion spectra in the decays 
$\Upsilon(mS)\to\Upsilon(nS)\pi\pi$ ($m>n, m=3,4,5, n=1,2,3$). This fact, first, 
confirms existence of the $f_0(1370)$ (up to now there is no firm conviction if it 
exists or not). Second, that exciting role of this meson in making the threshold 
bell-shaped form of the dipion spectra can be explained as follows: the $f_0(1370)$, 
being predominantly the $s{\bar s}$ state~\cite{SBLKN-prd14} and practically not 
contributing to the $\pi\pi$-scattering amplitude, influences noticeably the 
$K\overline{K}$ scattering; e.g., it was shown that the $K\overline{K}$-scattering 
length is very sensitive to whether this state does exist or not~\cite{SKN-epja02}. 
The interference of contributions of the $\pi\pi$-scattering amplitude and the 
analytically-continued $\pi\pi\to K\overline{K}$ and $\pi\pi\to\eta\eta$ amplitudes 
leads to the observed results.

It is important that we have performed a combined analysis of available data on 
the processes $\pi\pi\to\pi\pi,K\overline{K},\eta\eta$, on above-indicated decays 
of charmonia and bottomonia with the data from many collaborations. 
The convincing description (when including also the $\eta\eta$ channel) of 
the mentioned processes confirmed all our previous conclusions on the unified 
mechanism of formation of the basic dipion and $K\overline{K}$ spectra, 
which is based on our previous conclusions on wide 
resonances~\cite{SBLKN-jpgnpp14,SBLKN-prd14} and is related to contributions 
of the $\pi\pi$, $K\overline{K}$ and $\eta\eta$ coupled channels including 
their interference. This also confirmed all our earlier results 
on the scalar mesons \cite{SBLKN-prd14}; the most important results are:
\begin{enumerate}
\item Confirmation of the $f_0(500)$ with a mass of about 700~MeV and
a width of 930~MeV (the pole position on sheet~II is
$514.5\pm12.4-465.6\pm5.9$~MeV).
\item
An indication that the $f_0(980)$ (the pole on sheet~II is
$1008.1\pm3.1-i(32.0\pm1.5)$~MeV) is a non-$q{\bar q}$ state.
\item
An indication for the ${f_0}(1370)$ and $f_0 (1710)$ to have a dominant
$s{\bar s}$ component.
\item
An indication for the existence of two states in the 1500-MeV region: 
the $f_0(1500)$ ($m_{res}\approx1495$~MeV, $\Gamma_{tot}\approx124$~MeV) and
the $f_0^\prime(1500)$ ($m_{res}\approx1539$~MeV,
$\Gamma_{tot}\approx574$~MeV).

\end{enumerate}

\begin{acknowledgments}

This work was supported in part by the Heisenberg-Landau Program, 
by the Votruba-Blokhintsev Program for Cooperation of Czech Republic with JINR, 
by the Grant Agency of the Czech Republic (Grant No. P203/15/04301), 
by the Grant Program of Plenipotentiary of Slovak Republic at JINR, 
by the Bogoliubov-Infeld Program for Cooperation of Poland with JINR, 
by the BMBF (Project 05P2015, BMBF-FSP 202), by CONICYT (Chile) PIA/Basal FB0821, 
by Tomsk State University Competitiveness Improvement Program, 
by the Russian Federation program ``Nauka''(Contract No. 0.1764.GZB.2017), 
by Tomsk Polytechnic University Competitiveness Enhancement Program 
(grant No. VIU-FTI-72/2017), and by the Polish National Science Center (NCN) 
grant DEC-2013/09/B/ST2/04382.

\end{acknowledgments}

\appendix\section{}

The elements of $S$-matrix for 3-channel $\pi\pi$ scattering are represented via 
their moduli $\eta_{\alpha\alpha}=|S_{\alpha\alpha}|$, 
$\xi_{\alpha\beta}=|S_{\alpha\beta}|~~(\alpha\not=\beta)$ and 
phase shifts $\delta_{\alpha\alpha}$, 
$\phi_{\alpha\beta}$ ($\alpha,\beta=1,~2,~3$) as follows:
\begin{equation}\label{S_ph}
S_{\alpha\alpha}=\eta_{\alpha\alpha}e^{2i\delta_{\alpha\alpha}},~~~~~
S_{\alpha\beta}=i\xi_{\alpha\beta}e^{i\phi_{\alpha\beta}}~~~(\alpha\not=\beta).
\end{equation}
The phase shifts in Eq.~(\ref{S_ph}) have the form:
\begin{eqnarray}
&&\Phi_{11}(s) = \arg(S^{res}_{11}(s)) + \pi + 2B_1(s), 
~~B_1(s)=\sqrt{\frac{s-s_1}{s_1}}\Bigl(a_{11}+a_{1\sigma}
\frac{s-s_\sigma}{s_\sigma}\theta(s-s_\sigma)+
a_{1v}\frac{s-s_v}{s_v}\theta(s-s_v)\Bigr),\nonumber\\
&&\delta_{11}(s) = \Phi_{11}(s)\theta(M_1^2-s) + (\Phi_{11}(s) + 2\pi)
\theta(s-M_1^2)\theta(M_5^2-s) +(\Phi_{11}(s) + 4\pi)\theta(s-M_5^2),
\end{eqnarray}
\begin{eqnarray}
&&\Phi_{22}(s) = \arg(S^{res}_{22}(s)) + 2B_2(s), 
~~B_2(s)=\sqrt{\frac{s-s_2}{s_2}}\Bigl(a_{21}+a_{2\sigma}
\frac{s-s_\sigma}{s_\sigma}\theta(s-s_\sigma)+
a_{2v}\frac{s-s_v}{s_v}\theta(s-s_v)\Bigr),\nonumber\\
&&\delta_{22}(s) = \Phi_{22}(s)\theta(M_{12}^2-s) 
+ \Phi_{22}(s) - 2\pi\theta(s -M_{12}^2)\theta(M_{13}^2-s) +
\Phi_{22}(s)\theta(s-M_{13}^2)\theta(M_{14}^2-s) + \nonumber\\
&&(\Phi_{22}(s) - 2\pi)\theta(s-M_{14}^2),
\end{eqnarray}
\begin{eqnarray}
&&\Phi_{33}(s) = \arg(S^{res}_{33}(s) + 2B_3(s), 
~~B_3(s)=\sqrt{\frac{s-s_3}{s_3}}\Bigl(a_{31} + a_{1\sigma}
\frac{s-s_\sigma}{s_\sigma}\theta(s-s_\sigma) + a_{3v}
\frac{s-s_v}{s_v}\theta(s-s_v)\Bigr),\nonumber\\
&&\delta_{33}(s) = \Phi_{33}(s)\theta(M_{15}^2-s) 
+ (\Phi_{33}(s) + 2\pi)\theta(s-M_{15}^2),
\end{eqnarray}
\begin{eqnarray}
&&\Phi_{12}(s) = \arg(S^{res}_{12}(s)) + \pi + B_1(s)+ B_2(s),\nonumber\\
&&\phi_{12}(s) = \Phi_{12}(s)\theta(M_2^2-s) 
+ (\Phi_{12}(s) + \pi)\theta(s-M_2^2)\theta(M_3^2-s) 
+ (\Phi_{12}(s) + 2\pi)\theta(s-M_3^2),
\end{eqnarray}
\begin{eqnarray}
&&\Phi_{13}(s) = \arg(S^{res}_{13}(s)) + B_1(s)+ B_3(s),\nonumber\\
&&\phi_{13}(s) = \Phi_{13}(s)\theta(M_6^2-s) + (\Phi_{13}(s) + \pi)\theta(s-M_6^2),
\end{eqnarray}
\begin{eqnarray}
&&\Phi_{23}(s) = {\rm Re}[\arg(S^{res}_{23}(s)) 
+ B_2(s)+ B_3(s)] + \pi + 0.914406~\theta(s-s_3),\nonumber\\
&&\phi_{23}(s)=\Phi_{23}(s)\theta(M_7^2-s) 
+ (\Phi_{23}(s) + \pi)\theta(s-M_7^2)\theta(s_3-s) 
+ \Phi_{23}(s)\theta(s-s_3)\theta(M_8^2-s) +\nonumber\\
&& (\Phi_{23}(s) + \pi)\theta(s-M_8^2)\theta(M_9^2-s) 
+ (\Phi_{23}(s) - \pi)\theta(s-M_9^2)\theta(M_{10}^2-s) +\nonumber\\
&&\Phi_{23}(s)\theta(s-M_{10}^2)\theta(M_{11}^2-s) 
+ (\Phi_{23}(s) + \pi)\theta(s-M_{11}^2),
\end{eqnarray}
where ~$M_1=1.0003704$, ~$M_2=1.1699$, ~$M_3=1.4355$, ~$M_4=0.5620973$,  
~$M_5=1.6260221$, $M_6=1.4032403$, $M_7=1.0376107$, $M_8= 1.1946199$,  
$M_9=1.3788410$, $M_{10}=1.4356518$, $M_{11}=1.5349385$, 
$M_{12}=1.17160511$, $M_{13}=1.32724876$, $M_{14}=1.51411720$, $M_{15}=1.48837285$;
$a_{11}=0.0$, $a_{1\sigma}=0.0199$, $a_{1v}=0.0$, 
$a_{21}=-2.4649$, $a_{2\sigma}=-2.3222$, $a_{2v}=-6.611$, 
$a_{31}=-0.37755$, $a_{3\sigma}=0.8209$, $a_{3v}=-2.74575$; 
$s_\sigma=1.6338~{\rm GeV}^2$, $s_v=2.0857~{\rm GeV}^2$.
The values of energies $M_i$ and $M_{jk}$, at which the corresponding phases 
possess discontinuities corrected by adding an appropriate multiple of $90^\circ$, 
are determined empirically and relate to the used experimental data.

The resonance parts of $S$-matrix elements $S^{res}_{ij}(s)$ are parametrized 
by poles and zeros, representing resonances, with using the Le Couteur--Newton 
relations which in the 3-channel case and on the $w$-plane are shown 
in our work~\cite{SBLKN-jpgnpp14}. For calculation of moduli of the $S$-matrix 
elements there are needed also their inelastic background parts:
\begin{eqnarray}
&&S^{bgr}_{ii}=\exp\Bigl\{-2\theta(s-s_i)\sqrt{\frac{s-s_i}{s_i}}\Bigl(b_{i1}+
b_{i\sigma}\frac{s-s_\sigma}{s_\sigma}\theta(s-s_\sigma)+
b_{iv}\frac{s-s_v}{s_v}\theta(s-s_v)\Bigr)\Bigr\},\\
&&S^{bgr}_{ij}=\sqrt{S^{bgr}_{ii}S^{bgr}_{jj}}~~~(i<j),
\end{eqnarray}
where $b_{11}=b_{1\sigma}=0.0$, $b_{1v}=0.0338$, $b_{21}=b_{2\sigma}=0.0$, 
$b_{2v}=7.073$, $b_{31}=0.6421$, $b_{3\sigma}=0.4851$, $b_{3v}=0.0$.

\end{document}